 \documentclass[final,3p,times]{elsarticle}

\usepackage{amssymb}
\usepackage{amsmath}
\usepackage{bm} %for bold math
\usepackage{svg}
\usepackage{svg-extract}
\usepackage{booktabs}
\usepackage{multirow}
\usepackage{setspace} % For single/double spacing
\usepackage{lineno}   % For line numbers
\usepackage{siunitx}
\DeclareSIUnit{\molar}{M}
\biboptions{sort&compress} % To have [1-3]

\journal{Chemical Engineering Journal}

\begin{document}

% For review: double spacing and line numbers
%\doublespacing
%\setstretch{2.5}
%\linenumbers

\begin{frontmatter}

\title{Surface-access limitation in catalytic porous monoliths: Performance diagnosis using pore-resolved CFD}

\affiliation[inst1]{organization={CHAOS, Department of Chemical Engineering},%Department and Organization
            addressline={Polytechnique Montreal, PO Box 6079, Stn Centre-Ville}, 
            city={Montreal},
            postcode={H3C 3A7},
            state={Quebec},
            country={Canada}}
\affiliation[inst2]{organization={CREPEC, Department of Chemical Engineering},%Department and Organization
           addressline={Polytechnique Montreal, PO Box 6079, Stn Centre-Ville}, 
            city={Montreal},
            postcode={H3C 3A7},
            state={Quebec},
            country={Canada}}            
\affiliation[inst3]{organization={
%CREPEC, 
Department of Chemical and Biotechnological Engineering},%Department and Organization
            addressline={Université de Sherbrooke, 2500 Bd de l'Université}, 
            city={Sherbrooke},
            postcode={J1K 2R1},
            state={Quebec},
            country={Canada}}

\author[inst1]{Olivier Guévremont}
\author[inst2]{Olivier Gazil}
\author[inst3]{Federico Galli}
\author[inst2]{Nick Virgilio}
\cortext[cor1]{Corresponding author.}
\author[inst1]{Bruno Blais\corref{cor1}}      
\ead{bruno.blais@polymtl.ca}

\begin{abstract}
%%%%%%%%%%%%%%%%%% ABSTRACT

Porous  monoliths are promising catalyst supports due to their high surface area, interconnected channels, thermal stability and mechanical robustness. However, their tunable topology complicates design: trade-offs between conversion and pressure drop are not reliably captured by macroscopic descriptors, such as porosity, specific surface area, or tortuosity. Pore-resolved computational fluid dynamics~(PRCFD) addresses this gap by resolving pore-scale flow and transport, enabling diagnostics and discrimination between macroscopically similar structures.

We investigate surface-access-boundedness: a case where conversion is limited by flow maldistribution and incomplete utilisation of the catalytic surface, even at low Damköhler numbers~($\mathrm{Da}<1$). Using palladium-nanoparticle-coated silicone monoliths for \textit{p}-nitrophenol reduction, we perform reactive PRCFD in microcomputed-tomography-based geometries, calibrate a pseudo-heterogeneous eggshell reaction model, and validate transferability across samples and flow rates. 
We then diagnose surface-access-boundedness via the limited influence of diffusivity and reaction kinetics on conversion. 
Furthermore, we compare synthesised random monoliths with triply periodic minimal surface structures under matched porosity and surface area. 
Significantly, the required pumping power can decrease by up to an order of magnitude for the same molar production rate, depending on topology. These results show that, in heterogeneous systems affected by surface-access limitations, reactor performance is governed by structure-dependent surface accessibility rather than intrinsic kinetics or molecular diffusion alone, and that validated reactive PRCFD provides a practical framework to diagnose and compare porous reactor geometries under realistic operating conditions.

\end{abstract}

\begin{highlights}
    \item Synthesis of novel palladium-nanoparticles-covered silicone monoliths adapted to numerical study of heterogeneous catalysis.
    \item Design and calibration of a CFD model for reactive flows in porous media.
    \item Validation of model transferability across porous media using a volumetric representation of heterogeneous reaction.
    \item Demonstration of structure-induced surface-access limitations.
\end{highlights}


\end{frontmatter}

%%%%%%%% INTRODUCTION

\section{Pore-resolved computational fluid dynamics for reactive flows}
\label{sec:littrev}

    Most reactors in the chemical and fuel industries rely on heterogeneous catalysis~\cite{fogler2005elements}. Since this process occurs at the solid-fluid interface, porous catalysts are widely used to provide high surface areas or specific pore connectivity and size, which enhance access to active sites and can promote mixing~\cite{fogler2005elements,cybulski2005structured}. Among catalyst supports, porous monoliths are increasingly considered as alternatives to packed beds and other structured reactors~\cite{roy2004attritionMonoliths,kapteijn2022perspectivesMonoliths}. Their open, interconnected channel networks reduce pressure drop, mitigate dead zones and channelling, and promote uniform, plug-flow-like concentration profiles. When curvature and inertial effects become significant, their tortuous structures can induce recirculation and eddy formation, enhancing transverse transport and micro-mixing~\cite{chen2024inertiamixing,acharya2007RTDlimits,ganguli2024reviewMicromixers}. Compared to packed beds, the additional advantages in the fluid phase include the absence of particle contact points and the associated mass transfer resistance of these stagnant regions. In the solid phase, the absence of contact points also improves heat conduction and mechanical robustness, while reducing attrition and deformation~\cite{tomasic2006stateOfArtMonoliths,roy2004attritionMonoliths}. 
    Despite these advantages, the adoption of monolithic catalysts remains limited, in part because the links between structure, transport, and reactor performance are difficult to generalise~\cite{kapteijn2022perspectivesMonoliths,koufou2023scaling3Dprint,ladd2021RxPorousReviewProspects}. 
    In particular, the trade-offs between conversion, pressure drop, selectivity, and ultimately, molar production rate versus pumping power, remain difficult to reconcile, limiting full exploitation of the expanding design space enabled by versatile manufacturing techniques (e.g., subtractive and additive manufacturing)~\cite{gascon2015structuring,kapteijn2022perspectivesMonoliths}. 

    Heterogeneous reactors can be classified according to the rate-limiting step governing overall conversion~\cite{fogler2005elements,levenspiel1999}. In \textit{kinetics-limited regimes}, transport processes are sufficiently fast that reactant concentrations remain nearly uniform at the catalyst surface, and the overall reaction rates are controlled by intrinsic kinetics~\cite{fogler2005elements,levenspiel1999}. In \textit{diffusion-limited regimes}, molecular diffusion limits the transport of reactants to the active surface, leading to concentration gradients at the interface~\cite{fogler2005elements,levenspiel1999,froment1990chemical}. 
    These regimes can be framed using classical dimensionless numbers~\cite{fogler2005elements,levenspiel1999,froment1990chemical}. Fast heterogeneous reactions are characterised by large Damköhler numbers~(Da $\gg$ 1), which indicate that intrinsic reaction rates exceed characteristic transport rates. Depending on the transport process considered, Damköhler numbers compare intrinsic rates either to advective transport or to molecular diffusion, external or internal to the catalyst.
    The Thiele modulus is a more commonly used number that relates reaction kinetics to internal diffusion and can be used to estimate the effectiveness factor $\eta$, defined as the ratio of the observed reaction rate to the rate that would occur if the concentration throughout the solid were equal to the bulk concentration.
    Large Thiele moduli correspond to $\eta < 1$, indicating significant internal diffusion limitations~\cite{fogler2005elements,levenspiel1999,froment1990chemical}.
    Ideal reactors would exhibit infinitely fast diffusion and the system would correspond to a purely kinetics-limited regime~\cite{fogler2005elements,levenspiel1999}. 
    Real porous reactors, however, have finite diffusivity, complex pore geometries, and non-uniform flow paths that result in only a fraction of the surface being effectively accessed~\cite{whitaker1986darcy,quintard1993transport,ladd2021RxPorousReviewProspects,delgado2006reviewDispersion,golfier2002darcydissolution}. 
    The regimes outlined above fail to encompass \textit{surface-access‑boundedness}: a structure- and flow-organisation-driven limitation where only a fraction of the catalytic surface is effectively reached by the reactants. The important distinction here is that increasing molecular diffusivity alone does not remove this limitation, nor does changing the intrinsic kinetics~\cite{whitaker1986darcy,quintard1993transport,ladd2021RxPorousReviewProspects,delgado2006reviewDispersion}. This is not strictly a regime but a limiting mechanism that must be considered at the reactor scale. 
    This behaviour emerges only when pore-scale heterogeneities in flow and surface exposure are explicitly resolved, and cannot be diagnosed reliably using averaged transport descriptions.

    Despite the central role of transport and surface accessibility in fast-reaction regimes, porous reactor design is often guided by global geometric descriptors. Porosity, specific surface area, tortuosity, and characteristic pore size are common metrics for comparing and screening structures, as they provide compact measures of the available catalytic surface and its organisation~\cite{kapteijn2022perspectivesMonoliths,cybulski2005structured}. Using these metrics often involves the assumption that the catalytic surface is uniformly accessible and used by the reactants~\cite{whitaker2013VolumeAveraging,quintard1993VA_overCell}. In practice, they provide little information on flow distribution within the pore network, the presence of preferential channels or stagnant regions, the efficiency of near-wall replenishment, or the heterogeneity of residence times experienced by reactants, although residence time distribution~(RTD) analyses can quantify such heterogeneity in an averaged, reactor-scale sense~\cite{berard2020rtd,delgado2006reviewDispersion,acharya2007RTDlimits}. 
    Accordingly, reactors with similar porosity and specific surface area can exhibit markedly different conversions, selectivities, and pressure drops. In the absence of detailed information, reactor design has therefore traditionally relied on conservative assumptions and safety factors to ensure performance across a range of operating conditions. This approach has proven effective for robust operation, but it can mask structure-induced transport limitations that need to be resolved to enable performance optimization and comparison between competing reactor designs~\cite{ladd2021RxPorousReviewProspects,golfier2002darcydissolution}.

    Experimental studies of porous reactors are costly and most commonly provide access to integral observables, such as global conversion, pressure drop, or RTD moments, without direct information on pore-scale flow organisation or local surface utilisation~\cite{levenspiel1999,delgado2006reviewDispersion}. Reduced-order or volume-averaged models, such as one-dimensional plug-flow~(PFM) or axial-dispersion models~(ADM)~\cite{danckwerts1953RTD}, volume-averaged Navier-Stokes~(VANS) formulations~\cite{quintard1993VA_overCell,whitaker2013VolumeAveraging,geitani2023vans}, RTD-based approaches, and pore network models~(PNM)~\cite{blunt2013poreScaleImaging}, rely on spatial averaging or simplified representations of the pore space. 
    These models are computationally efficient and valuable for reactor-scale analysis, but they rely on assumptions that filter pore-scale heterogeneities in velocity, concentration, and surface accessibility. 
    PFR, ADM and RTD-based approaches assume transverse homogeneity, and all these reduced-order models assume uniform access to the surface within an averaging volume. Furthermore, they rely on closure relations or empirical correlations (e.g., Sherwood-Reynolds-Schmidt correlations) to model the effects of mass transfer and mixing~\cite{froment1990chemical,whitaker2013VolumeAveraging}. Transport and reactions are therefore described in terms of effective parameters or global effectiveness factors, with little to no access to the spatial organisation of flow and reactions at the pore scale~\cite{froment1990chemical,delgado2006reviewDispersion}. These assumptions are problematic in fast-reaction regimes, where conversion depends on the continuous renewal of reactants at reactive interfaces. Reduced-order models capture reactor-scale trends and ensure robust operation, but they do not resolve structure-induced channelling, stagnant regions, or near-wall transport limitations~\cite{acharya2007RTDlimits,golfier2002darcydissolution}. More fundamentally, they do not provide access to local wall fluxes, the fraction of catalytic surface that is truly used, or the distribution of reaction rates within the pore network~\cite{acharya2007RTDlimits,ladd2021RxPorousReviewProspects}. 

    Pore-resolved computational fluid dynamics~(PRCFD) is a physically grounded alternative where pore-scale structure and flow field are explicitly resolved.  
    PRCFD enables the prediction and localisation of structure-induced hydrodynamic anomalies that are smoothed out by volume-averaged descriptions~\cite{flaischlen2023structureHydrodynamicsRTDCFD}.
    Related approaches, such as lattice Boltzmann methods, have been used to investigate flow and transport in complex porous structures, typically relying on simplified formulations for reactive processes~\cite{blunt2013poreScaleImaging,belot2021b}. 
    Historically, application of PRCFD to porous reactors was limited by prohibitive computational cost, scarcity of experimentally supported datasets at the pore scale~\cite{santos2022dataset,noiriel2021pore_review}, and uncertainty in specifying transferable surface-reaction descriptions; kinetic parameters are frequently regressed from limited data and can be non-unique~\cite{jurtz2019review}. 
    Recent advances in high-performance computing and imaging now enable credible coupled studies. 
    Boigné et al.~\cite{boigne2024iRmicroCT} combined synchrotron microcomputed tomography~(\textmu CT), infrared thermography, and pore-resolved simulations to analyse porous media combustion, illustrating how spatial diagnostics can be used to benchmark pore-scale models.
    Dong et al.~\cite{dong2018rxPRCFD} validated \textmu CT-based reactive PRCFD for CO oxidation in platinum-coated ceramic foams and identified mass-transport limitations, while Maes and Menke~\cite{maes2021geochemfoam} verified an open-source OpenFOAM-based multiphase reactive-transport workflow on canonical benchmarks and demonstrated it on \textmu CT geometries. Consistent with these developments, Acharya et al.~\cite{acharya2007RTDlimits} showed that pore-scale resolution may be required to predict RTDs and performance when mixing/diffusion limitations dominate.
    Together, these studies establish PRCFD as a powerful tool for investigating transport-reaction coupling in porous structures, with growing experimental support.
    In practice, the choice between reduced-order models and PRCFD is a trade-off between computational cost and the level of physical detail required for a given analysis~\cite{fogler2005elements}. Prior PRCFD studies have largely focused on predicting performance or identifying transport limitations in specific systems, rather than on systematically diagnosing structure-dependent surface accessibility or enabling fair comparisons across macroscopically similar reactor configurations.
    %metric-consistent comparisons across different reactor geometries.

    In this work, we extend our previously developed PRCFD framework based on the finite element method~(FEM)~\cite{guevremont2024poreresolvedcfddigitaltwin} to reactive flows in \textmu CT-based digital twins of palladium-nanoparticle-coated silicone monoliths, then calibrate and validate a pseudo-heterogeneous reaction model against experimental conversion data across multiple samples and Reynolds numbers. The results show that conversion is governed primarily by structure-dependent surface accessibility rather than by intrinsic kinetics or molecular diffusivity, thereby identifying surface-access-boundedness as the dominant limitation and establishing reactive PRCFD as a tool to diagnose and compare porous monolith geometries using conversion, pressure drop, and related process-intensification metrics.
    \textit{p}-nitrophenol reduction over palladium nanoparticles~(PdNPs) is used here as a representative fast, surface-limited reaction system, but the proposed approach is applicable to other heterogeneous processes operating under similar conditions: liquid-phase hydrogenations~\cite{fogler2005elements,froment1990chemical,cybulski2005structured}, gas-phase oxidation reactions on structured supports~\cite{dong2018ctcfd_fixedbedreactor_ht_species}, or surface-limited biofilm-mediated reactions, where transport to reactive interfaces controls overall performance~\cite{rittmann2001biofilm_surfacelimit}. The scope of this work is limited to the diagnosis and screening of structure-induced transport limitations. Its central contribution is to demonstrate surface-access-boundedness using experimentally validated reactive PRCFD and to show how this limitation can be diagnosed and mitigated across different porous geometries under similar conditions.

\section{Methods and framework: reduction of \textit{p}-nitrophenol on Pd/silicone monoliths}

    \subsection{Reactive PRCFD framework}
    
        \subsubsection{System, chemistry and kinetics model}
        
            We consider the reduction of \textit{p}-nitrophenol catalysed by PdNPs supported on porous silicone monoliths, hereafter denoted PdNPs@silicone monoliths. 
            This PdNP@silicone system is used as a model catalyst to isolate surface-access effects under flow-through conditions: the activation protocol (described in \textbf{Section~\ref{sec:sample_preparation}}) is designed to synthesise PdNPs at the solid-fluid interface, which motivates defining reactions in a thin interfacial shell using the signed distance function~(SDF). 
            
            At low \textit{p}-nitrophenol concentrations and with an excess of reducing agent (here sodium borohydride), the reaction can be simplified to a first‑order reaction depending solely on the \textit{p}‑nitrophenol concentration, with isothermal conditions~\cite{gazil2023urethane1storder,chatterjee2021reduction1storder}. 
            We define the reaction rate as the volumetric sink term $S_c$, where $k$~$[\SI{}{\per\second}]$ is the reaction constant and $c$ is the reactant concentration.
            
            \begin{align} 
                S_c = - k c
                \label{eq:reaction_rate}
            \end{align}
            
            We model the heterogeneous (surface) reaction using a localised volumetric source term defined from the SDF, $\lambda$, to form a reactive shell near the solid-fluid interface~\cite{russo2020eggshellmodel}.
            We define the reaction constant through a shell thickness parameter $\varepsilon_k$ and a peak value $k_0$ at the interface using a smooth, symmetric function that decays rapidly away from the interface.

            \begin{align} 
                k(\lambda) = k_0 \exp \left( - \frac{\lambda^2}{\varepsilon_k^2}\right)
                \label{eq:sdf_dependent_reaction}
            \end{align}
        
            We describe the system conditions using the Reynolds~(${\mathrm{Re}}$), Péclet~(${\mathrm{Pe}}$), and Damköhler~($\mathrm{Da}$) numbers:
        
            \begin{align}
                \mathrm{Re} = \frac{u_\mathrm{D} d_\mathrm{p}}{\nu }
                = \frac{Q d_\mathrm{p}}{A \varepsilon \nu }
                \label{eq:reynolds}
            \end{align}
        
            \begin{align}
            \mathrm{Pe} = \frac{u_\mathrm{D} d_\mathrm{p}}{\mathcal{D}} = \frac{Q d_\mathrm{p}}{A \varepsilon \mathcal{D}}
                \label{eq:peclet}
            \end{align}
            
            \begin{align}
                \mathrm{Da} = \frac{\tau_a}{\tau_r}=\left( \frac{A}{V} \int_{-\infty}^{+\infty} k(\lambda) d\lambda \right) \left(\frac{V\varepsilon}{Q} \right) = \frac{\sqrt{\pi}k_0 \varepsilon_k A \varepsilon }{Q}
                \label{eq:damkohler}
            \end{align}
                
            where $u_\mathrm{D}$ is the Darcy velocity~($=Q / \left( A \varepsilon \right)$)~\cite{wood2020review_turbFlo_porousmedia}, $Q$ the volumetric flow rate, $A$ the cross-section area, $d_\mathrm{p}$ the pore diameter, $\varepsilon$ the porosity, $\nu$ the kinematic viscosity, $\mathcal{D}$ the diffusivity, $\tau_r$ the reaction characteristic time, $\tau_a$ the residence time, and $V$ the cylinder volume. We approximate the reaction time scale by integrating the Gaussian formulation $k(\lambda)$ across the signed distance coordinates $\lambda$. The bounds $\left( -\infty,+\infty\right)$ provide the analytical integral of the localised reactive layer.
    
        \subsubsection{Governing equations}
            \label{sec:governing_equations}
        
            We obtain the pressure and velocity fields from the numerical resolution (\textbf{Section~\ref{sec:numerical_model_prcfd}}) of the transient incompressible Navier-Stokes equations, where $\bm{u}$ is the velocity, $p$ is the pressure, $\rho$ is the density, $\bm{\tau}$ is the deviatoric stress tensor, $f$ is a source term and $\nu$ is the kinematic viscosity. We include the density in the pressure definition such that $p^* = \frac{p}{\rho}$.
            \begin{flalign}
                \nabla \cdot \bm{u} &= 0  \label{eq:continuity}\\
                \frac{\partial \bm{u}}{\partial t} + (\bm{u} \cdot \nabla) \bm{u} &= - \nabla p^* + \nabla \cdot \bm{\tau} + \bm{f} \label{eq:transient_NS} \\
                \bm{\tau} &= \nu \left( (\nabla \bm{u}) + (\nabla \bm{u})^T \right)
                \label{eq:stress_tensor}
            \end{flalign}
            
            We use a single reactant concentration to describe the transport and reaction. Its solution is obtained by solving the transient advection-diffusion-reaction equation, where $S_c$ is the source term (reaction rate). 
            
            \begin{align} 
                \frac{\partial c}{\partial t} + \bm{u} \cdot \nabla c = \mathcal{D} \nabla^2 c + S_c 
                \label{eq:adv_dif_react}
            \end{align}
            
            The velocity field solution is transferred from the resolution of the Navier-Stokes equations in steady-state. Similarly to the reaction constant, the diffusivity is a function of the SDF and depends on a thickness parameter $\varepsilon_{\mathcal{D}}$ as well as inside and outside values $\mathcal{D}_-$ and $\mathcal{D}_+$
            
            \begin{align} 
                \mathcal{D}(\lambda) = \mathcal{D}_- + \left(\mathcal{D}_+ - \mathcal{D}_-\right) \left( 0.5 + 0.5 \tanh \left(\frac{\lambda}{\varepsilon_{\mathcal{D}}}\right)\right)
                \label{eq:sdf_dependent_diffusivity}
            \end{align}

            We set $\mathcal{D}_- \ll \mathcal{D}_+$, consistent with modelling the silicone phase as impermeable to the reactant over the experimental time scale. $\varepsilon_{\mathcal{D}}$ controls the thickness of the interfacial transition.

        \subsubsection{Numerical implementation}
            \label{sec:numerical_model_prcfd}
        
            We used the Finite Element Method~(FEM) to solve the weak form of the governing equations using the open-source software Lethe~\cite{alphonius2025lethe} (v1.0.3), built upon the deal.II open-source library~\cite{dealii2025}. The solver uses a streamline upwind Petrov-Galerkin and pressure-stabilizing Petrov-Galerkin~(SUPG/PSPG) stabilised finite element formulation to solve the Navier-Stokes equations, which allows the use of the same order of elements for velocity and pressure~\cite{blais2020lethe}. 
            We used hexahedral linear elements~(Q$_1$) for velocity, pressure and concentration~(Q$_1$Q$_1$Q$_1$). Transient simulations were performed using a backward difference formula of first order~(BDF1)~\cite{hay2015bdf_timeintegration}. Reactive simulations were solved in two parts: first, the flow field was established without reaction until steady-state; second, the flow field was frozen and we used larger time steps to resolve only the concentration field, until a new steady-state was reached. 
            Thus, steady-state solutions could be reached despite the high non-linearity of the problem.
            The separation in two parts with distinct time steps reduced drastically the computing costs. More details on the PRCFD numerical model and implementation are available on Lethe's GitHub repository\footnote{https://github.com/chaos-polymtl/lethe} as well as in previous publications~\cite{blais2020lethe, alphonius2025lethe, guevremont2024poreresolvedcfddigitaltwin}. 
            
    \subsection{Catalyst preparation}
        \label{sec:sample_preparation}
        
        We synthesised porous silicone monoliths following the procedure described in previous work~\cite{guevremont2024poreresolvedcfddigitaltwin,esquirol2014cocontinuousPolymers} using initial co-continuous blends of polystyrene~(PS) and polylactide~(PLA) annealed for \SI{45} and \SI{60}{\minute} and polymer proportions that lead to a final porosity of approximately 50\% in the silicone monoliths. 
        We activated silicone monoliths by synthesising PdNPs at the surface of the material, based on the method by Gazil et al.~\cite{gazil2020npPalladium}:

        \begin{enumerate}
            \item Submerge the samples in a \SI{6}{\milli\molar} palladium~(II) acetate (product 520764-1G by Aldrich Chemistry) solution in chloroform (product C607-4 by Thermo Fischer Scientific) for \SI{15}{\minute}. 
            \item Recover and carefully dry the monolith using low-lint disposable wipes, ensuring complete solvent removal from the pores.
            \item Submerge the samples in a \SI{30}{\milli\molar} sodium borohydride (product 71320-25G by Aldrich Chemistry) solution in water for \SI{5}{\minute}. Apply vacuum to ensure that hydrogen gas byproducts do not obstruct the pores. 
            \item Submerge the samples in distilled water for \SI{7}{\minute} to stop the reduction and remove the unreacted species. 
            \item Leave the chloroform swelled samples for \SI{2}{\hour} under a fume hood; the chloroform evaporation leads to the recovery of the samples' initial size.
            \item Encapsulate the samples in custom-made acrylic tubes sealed with dichloromethane. Each capsule is composed of three acrylic tubes: one of inner diameter~(ID) \SI{88}{\milli\metre}, and two of outer diameter~(OD) \SI{88}{\milli\metre} and ID \SI{62}{\milli\metre}. Place each sample in a large tube, immobilise the sample by inserting two small tubes to form a cavity of height \SI{60}{\milli\metre}, then fuse the tubes together using dichloromethane. 
        \end{enumerate}

\begin{figure}[htpb!]
            \centering
            \includegraphics[width=0.6\textwidth]{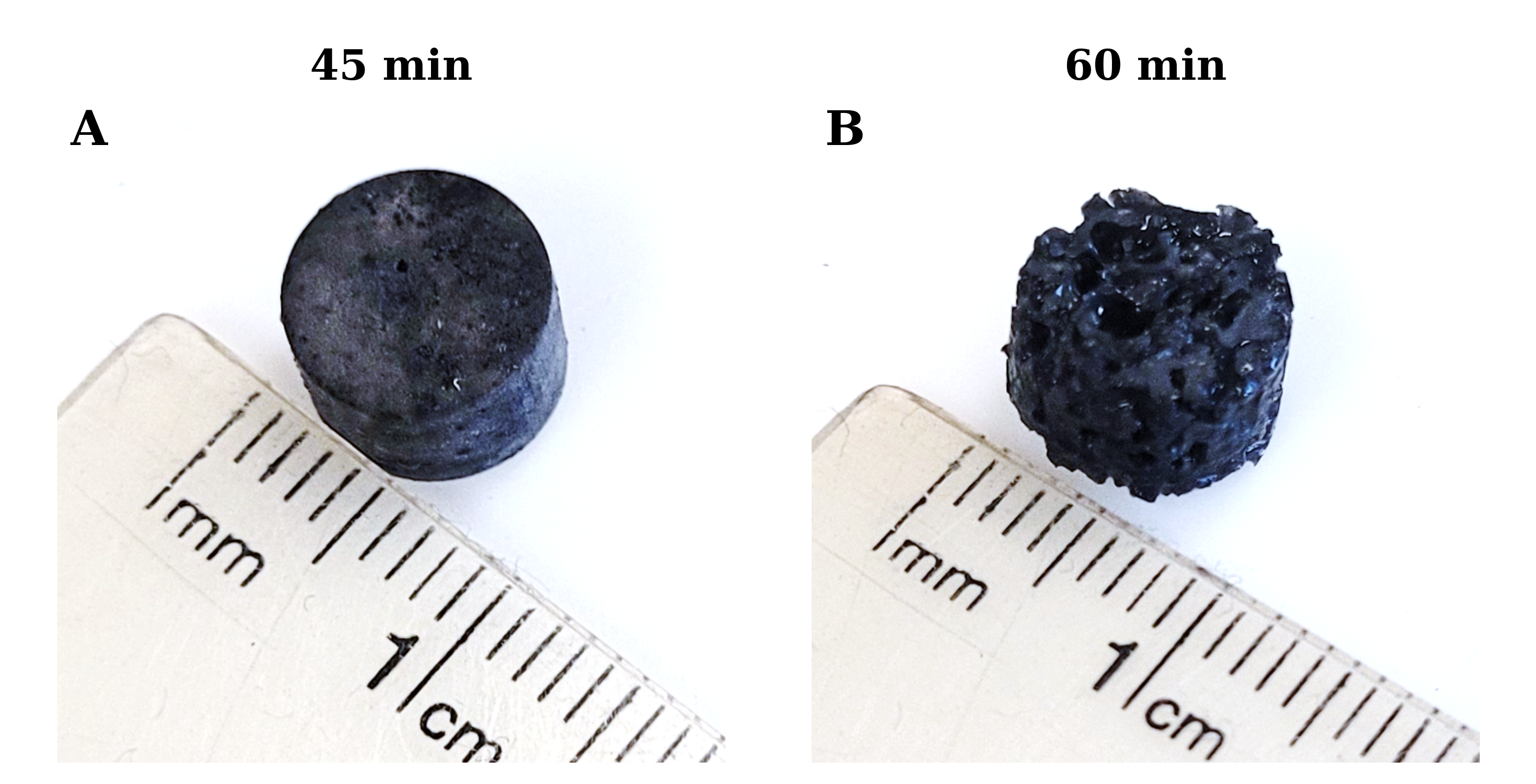}
            \caption{
            Monoliths after PdNPs synthesis. The \SI{45}{\minute}~(A) and \SI{60}{\minute}~(B) monoliths differ in annealing duration prior to silicone injection.
            }
            \label{fig:composite_with_ruler}
        \end{figure}

        The capsules are designed to mimic porous reactors design. 
        The cavity in each capsule is a cylinder (height: \SI{60}{\milli\metre}, diameter: \SI{88}{\milli\metre}) with circular inlet/outlet at the bases (diameter: \SI{62}{\milli\metre}). 
        \textbf{Figure~\ref{fig:composite_with_ruler}} illustrates the state of monoliths loaded with PdNPs after completion of step 5. 
        \textbf{\ref{sec:appendix_capsule_visualisation}} shows the capsule geometry once assembled.

    \subsection{Material characterisation}
        \label{sec:material_characterisation}
        
        \subsubsection{Porous structure characterisation}
\begin{figure}[htpb!]
                \centering
                \includegraphics[width=0.6\textwidth]{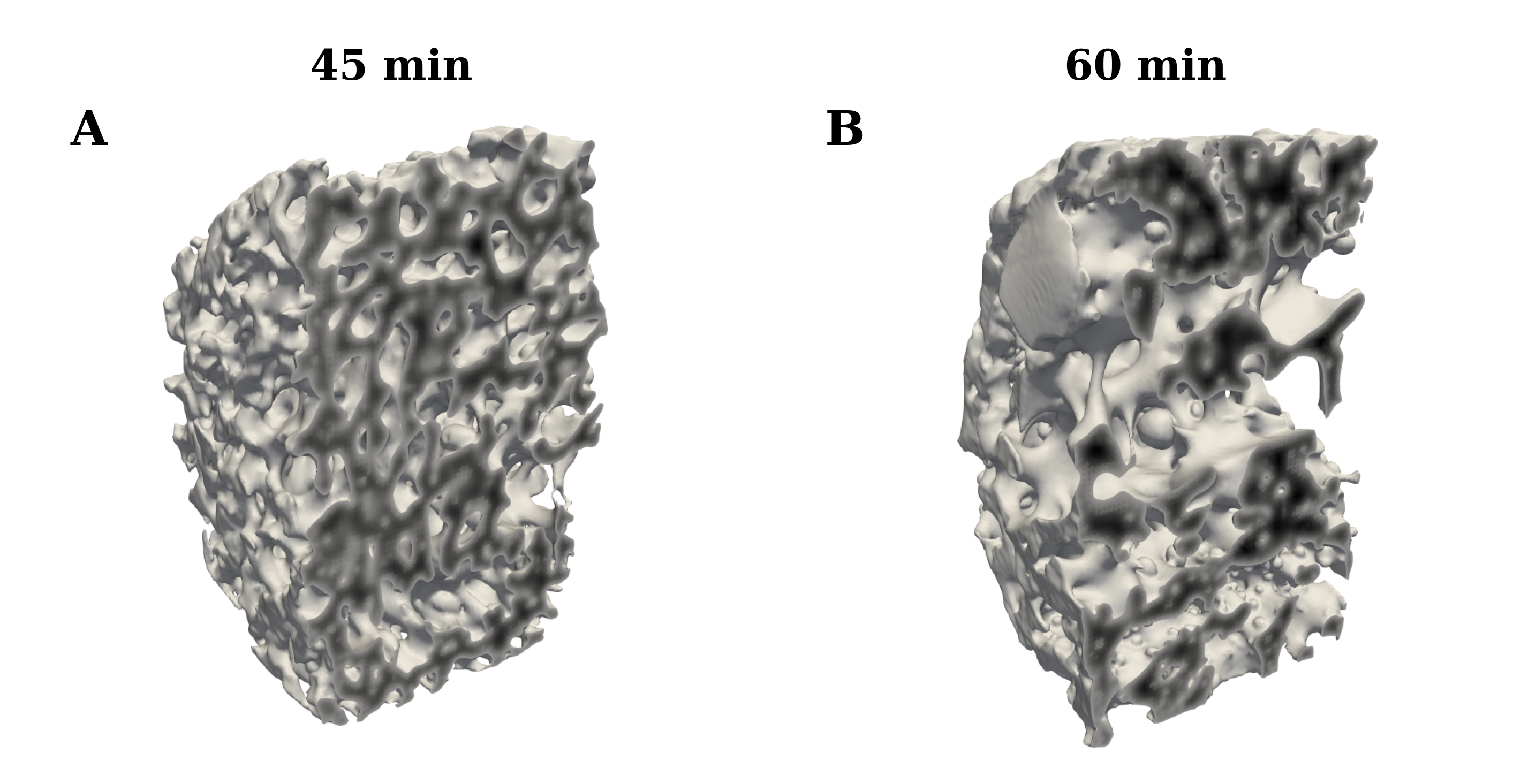}
                \caption{Digitised samples of \SI{45}{\minute}~(A) and \SI{60}{\minute}~(B). The pores are larger for longer annealing durations.}
                \label{fig:composite_CT}
            \end{figure}

            We digitised encapsulated samples using \textmu CT analysis with a ZEISS Xradia 520 Versa XRM instrument following the same procedure as our previous work~\cite{guevremont2024poreresolvedcfddigitaltwin}. We obtained each sample's unique structure from visual segmentation of the void and solid phases using Dragonfly Pro Version 2020.1, Build 809 (Object Research Systems, Inc, Montreal, Quebec, Canada). 
            Voxel sizes were set to be at most \SI{10}{\micro\metre}. Surface grids of the monoliths were generated at an intensity threshold of 50\%. \textbf{Figure~\ref{fig:composite_CT}} shows the inside structure obtained from \textmu CT analysis.

            We used Dragonfly's integrated workflow based on OpenPNM and its implementation of the Sub-Network of an Oversegmented Watershed algorithm~\cite{gostick2016openpnm,gostick2019porespy,gostick2017openpnm_watershed} to generate pore network models for each sample's void space. The distribution of inscribed pore diameters and the distance between each pair of connected pores (at centroid) are approximated as normal and log-normal, respectively. Mean values and standard deviations for pore diameter, $d_p$ and $\sigma_p$, and distance between connected pores, $d_c$ and $\sigma_c$, are shown in \textbf{Table~\ref{tab:samples_characterisation}}. Table~\ref{tab:samples_characterisation} also shows each solid's surface $S$, corrected by a $2/3$ factor to account for the staircase effect of voxelization~\cite{yeong1998a,yeong1998b}, volume $V$, corresponding specific surface $S_p$ and porosity $\varepsilon$ (based on the cylindrical capsule of diameter \SI{0.88}{\centi\metre} and height \SI{0.60}{\centi\metre}). Additional details on the porous structure are shown in \textbf{\ref{sec:appendix_pore_size_distributions}}.
            
            \begin{table}[htpb!]
                \centering
                \caption{
                Porous media descriptors for the PdNPs@silicone monoliths used in the reactive flow experiments.}
                \label{tab:samples_characterisation}
                \begin{tabular}{c c c c c c c c c}
                    \hline
                    \textbf{Identification} &  $d_p$ [\SI{}{\micro\metre}] & $\sigma_p$ [\SI{}{\micro\metre}]  &  $d_c$ [\SI{}{\micro\metre}] & $\sigma_c$ [\SI{}{\micro\metre}] & $S$ [\SI{}{\centi\metre\squared}] & $V$ [\SI{}{\centi\metre\cubed}]  & $S_p$ [\SI{}{\per\centi\metre}] & $\varepsilon$ [\SI{}{\percent}] \\
                    \cmidrule(lr){1-9}
                      \SI{45}{\minute}~\#1 & 400 & 200 & 400 & 200 & 23 & 0.20 & 115 & 45 \\ 
                      \SI{45}{\minute}~\#2 & 300 & 100 & 400 & 200 & 17  & 0.22 & 79 & 40 \\ 
                      \SI{45}{\minute}~\#3 & 500 & 200 & 500 & 200 & 10  & 0.20 & 52 & 45 \\ 
                      \SI{60}{\minute}~\#1 & 600 & 200 & 500 & 300 & 8  & 0.17 & 45 & 53 \\ 
                      \SI{60}{\minute}~\#2 & 700 & 200 & 600 & 300 & 8  & 0.19 & 41 & 49 \\ 
                      \SI{60}{\minute}~\#3 & 700 & 300 & 600 & 300 & 7  & 0.20 & 38 & 46 \\ 
                      %\SI{45}{\minute}~\#1 & 385 & 159 & 401 & 195 & 115 & 36 \\ 
                      %\SI{45}{\minute}~\#2 & 315 & 121 & 376 & 164 & 79  & 33 \\ 
                      %\SI{45}{\minute}~\#3 & 498 & 170 & 507 & 211 & 52  & 48 \\ 
                      %\SI{60}{\minute}~\#1 & 616 & 232 & 548 & 297 & 45  & 49 \\ 
                      %\SI{60}{\minute}~\#2 & 660 & 243 & 572 & 299 & 41  & 48 \\ 
                      %\SI{60}{\minute}~\#3 & 703 & 271 & 597 & 318 & 38  & 50 \\ 
                    \hline
                \end{tabular}
            \end{table}

        \subsubsection{Nanoparticles characterisation}

\begin{figure}[htpb!]
                \centering
                    \includegraphics[width=0.7\textwidth]{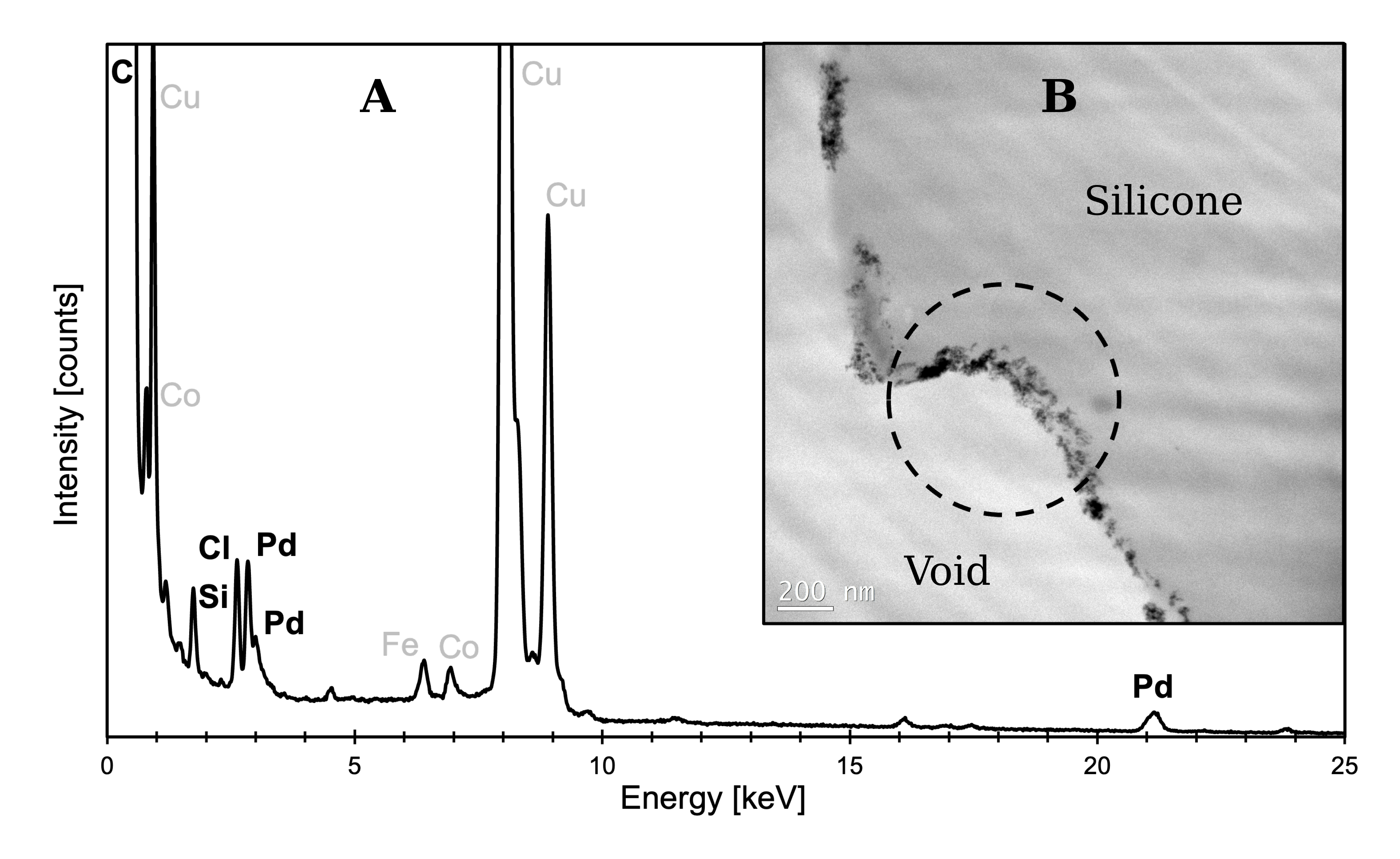}
                \caption{EDS spectrum analysis~(A) of circled region of the surface of PdNP@silicone~(B). The image corresponds to the interface of the sample \SI{60}{\minute}~\#3, obtained by TEM analysis.}
                \label{fig:EDS_and_frontiere}
            \end{figure}

            We obtained PdNPs composition, size, circularity and positioning from Transmission Electron Microscopy~(TEM) and Energy Dispersive X-ray Spectroscopy~(EDS). 
            TEM analyses were carried out using a MET-JEOL JEM-F200 cold-FEG operated at 200 kV in conventional TEM mode. The EDS spectrum was captured using a JEOL EDS spectrometer.
            \textbf{Figure~\ref{fig:EDS_and_frontiere}A} confirms the presence of palladium in the silicone support, as evidenced by the presence of the Pd peaks around \SI{2.8}{} and \SI{21}{\kilo\volt}, and the Si and Cl peaks at \SI{1.7}{} and \SI{2.6}{\kilo\volt}, respectively. The observed localisation of Pd at the interface (\textbf{Figure~\ref{fig:EDS_and_frontiere}B}) supports the use of an egg-shell model to prescribe interfacial reactivity in Equation~\ref{eq:sdf_dependent_reaction}~\cite{russo2020eggshellmodel}. 

            Additional analyses of clusters of PdNPs for both categories of PdNP@silicone monoliths reveal that nanoparticles have diameters below \SI{10}{\nano\metre} (with standard deviations of $3$) and uniform circularities \textit{ca.} $0.9$, as shown in \textbf{Table~\ref{tab:np_mean_characteristics}}. We provide details on the results and methodology in \textbf{\ref{sec:tem_pdnp}}.

            \begin{table}[htpb!]
                \centering
                \caption{PdNPs diameter and circularity mean and standard deviation for samples \SI{45}{\minute} and \SI{60}{\minute}.}
                \label{tab:np_mean_characteristics}
                \begin{tabular}{c c c }
                    \toprule
                    \multirow{1}{*}{\textbf{Identification}} 
                    & \multicolumn{1}{c}{\textbf{Diameter}~[\SI{}{\nano\metre}]} 
                    & \multicolumn{1}{c}{\textbf{Circularity}} \\
                    \cmidrule(lr){1-3}
                    \SI{45}{\minute} & $8 \pm 3$  & $0.89 \pm 0.03$  \\
                    \SI{60}{\minute} & $9 \pm 3$  & $0.88 \pm 0.04$  \\
                    \bottomrule
                \end{tabular}
            \end{table}

    \subsection{Reactive flow experiments}
        \subsubsection{\textit{p}-nitrophenol reduction protocol}
            \label{sec:experiment_flow_reduction}
            We ran experiments at flow rates in a range of Reynolds numbers from $0.25$ to $3.7$ using the experimental setup presented in \textbf{\ref{sec:experimental_setup}}. Each experiment was run three times. Gas produced by the reaction was removed by slight vibration throughout each experiment.
            Catalyst deactivation is assumed to be negligible. 
            The reactants mixture is composed of \textit{p}-nitrophenol (limiting reactant) and NaBH$_4$ (in excess). The diffusivity of \textit{p}-nitrophenol is estimated at $6.85\cdot 10^{-6}$\SI{}{\centi\metre\squared\per\second}~\cite{seida2022pnitrophenolDiffusivit,wilke1955correlationDiff}. NaBH$_4$ is in excess to ensure that the reaction is of first order with respect to \textit{p}-nitrophenol, as well as to maintain the steady effect of NaBH$_4$ on the reaction rate despite eventual parasitic reactions~\cite{gazil2023urethane1storder}. Moreover, this ensures that NaBH$_4$ concentration does not have to be measured nor tracked.
            We measured the concentration of upstream and downstream reactants using UV-vis. The resulting concentrations were used to compute the reactant conversion.

        \subsubsection{Conversion experimental results}
            \label{sec:conversion_results}
\begin{figure}[htpb!]
                \centering
                \includegraphics[width=0.8\textwidth]{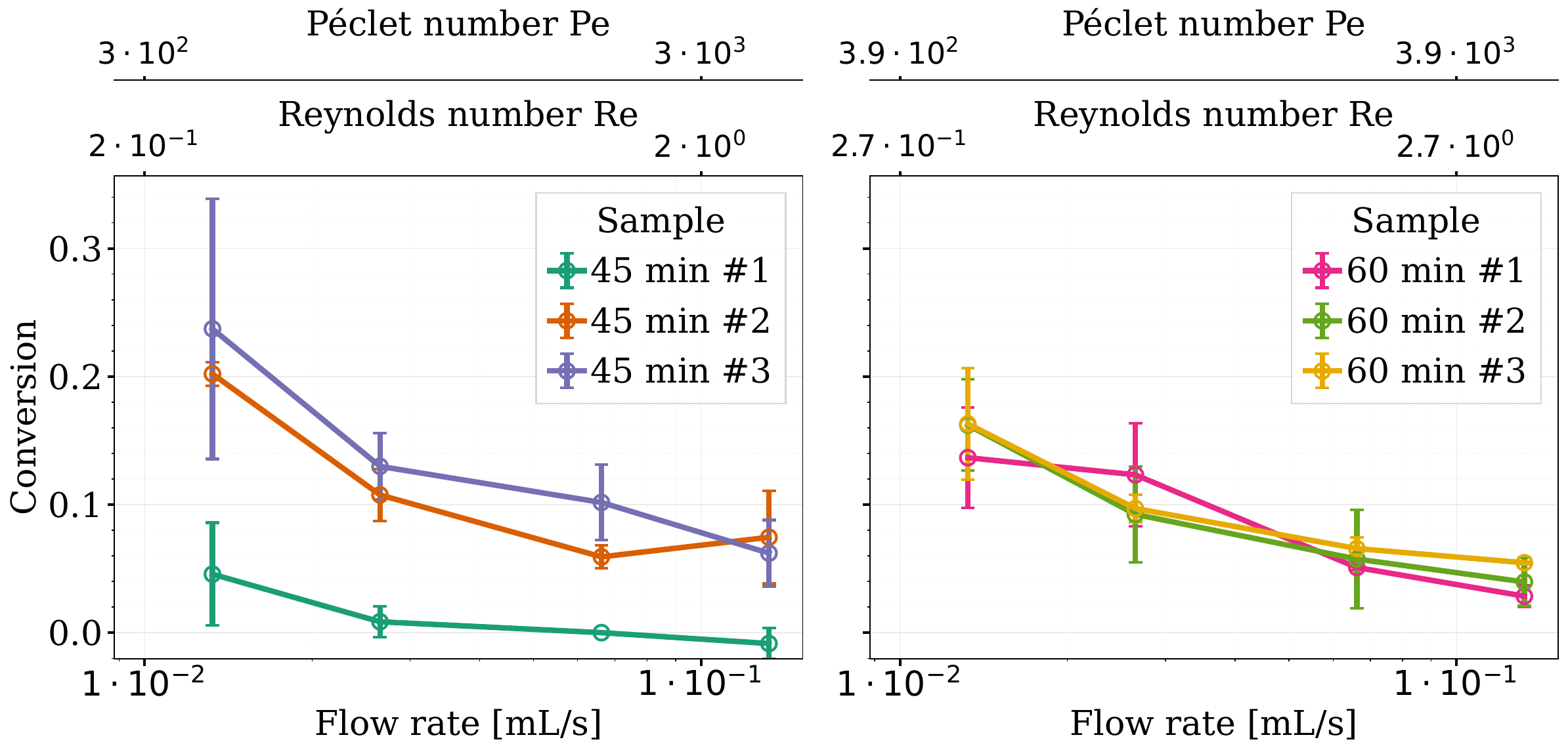}
                \caption{Conversion of \textit{p}-nitrophenol after flow-through PdNP@silicone monoliths, measured by UV-vis. 
                Plotted values represent the mean, and error bars (standard deviation) are calculated from three data points.
                } 
                \label{fig:experimental_conversion}
            \end{figure}

            The conversion results of the \textit{p}-nitrophenol reduction experiments are shown in \textbf{Figure~\ref{fig:experimental_conversion}}. We show the mean $\mu$ from $n=3$ experimental repetitions, for each flow rate and sample. The error bars are the standard deviation.
        
\section{Calibration and validation}
    \label{sec:calibration_transferability}
    In this section, we focus on the calibration of the reactive parameters followed by validation of the model through transferability to other samples, ensuring its predictivity.

    \subsection{Simulation setup}
        We simulated each case by running two distinct steps: (1)~flow simulation without reactant, until steady-state, (2)~reaction simulation with fixed velocity solution, until steady-state of the concentration field. 
        The steady-state criteria are that (1) pressure drop $\Delta p$ varies by $\Delta_\mathrm{\Delta p}<0.25\%$ and (2) conversion $X$ varies by $\Delta_\mathrm{X}<0.1\%$ between time steps.

        \begin{align} 
            \Delta_{\{\Delta p,X\}} = \frac{{\{\Delta p,X\}_{t_\mathrm{end}}}-{\{\Delta p,X\}_{t_\mathrm{end}-\Delta t}}}{{\{\Delta p,X\}_{t_\mathrm{end}}}} 
            \label{eq:relative_difference}
        \end{align}

        Simulation control parameters for both steps are shown in \textbf{Table~\ref{tab:lethe_simulation_control_parameters}}. Physical properties are shown in \textbf{Table~\ref{tab:lethe_physical_properties}}. We covered combinations of $k_0 \in \{1\cdot 10^{-2},\,3\cdot 10^{-2},\,1\cdot 10^{-1},\,3\cdot 10^{-1},\,1\}$ and $\varepsilon_k \in \{1.25\cdot 10^{-3},\,2.5\cdot 10^{-3},\,5\cdot 10^{-3},\,1\cdot 10^{-2},\,2\cdot 10^{-2}\}$ to calibrate the egg-shell reaction model. Grid convergence considerations and results are covered in \textbf{\ref{sec:grid_convergence_annex}}. The diffusivity transition thickness $\varepsilon_\mathcal{D}$ was set approximately equal to the minimum cell length, to regularise the diffusivity jump at the interface over one cell. The same criterion was used to define the minimal value of $\varepsilon_k$.

        We used smooth polynomial functions at the inlet boundary $\Gamma_\mathrm{in}$ to impose the volumetric flow rate while respecting no-slip boundary conditions at the adjacent walls.

        \begin{equation}
            \begin{aligned}
                Q &= \int_{\Gamma_\mathrm{in}} u_x \, \mathrm{d}\Gamma, \\
                u_x(y=\pm 0.6) &= 0, \\
                u_x(z=\pm 0.6) &= 0 .
            \end{aligned}
            \label{eq:imposed_velocity_profile}
        \end{equation}
        
\begin{figure}[htpb!]
            \centering
            \includegraphics[width=0.75\textwidth]{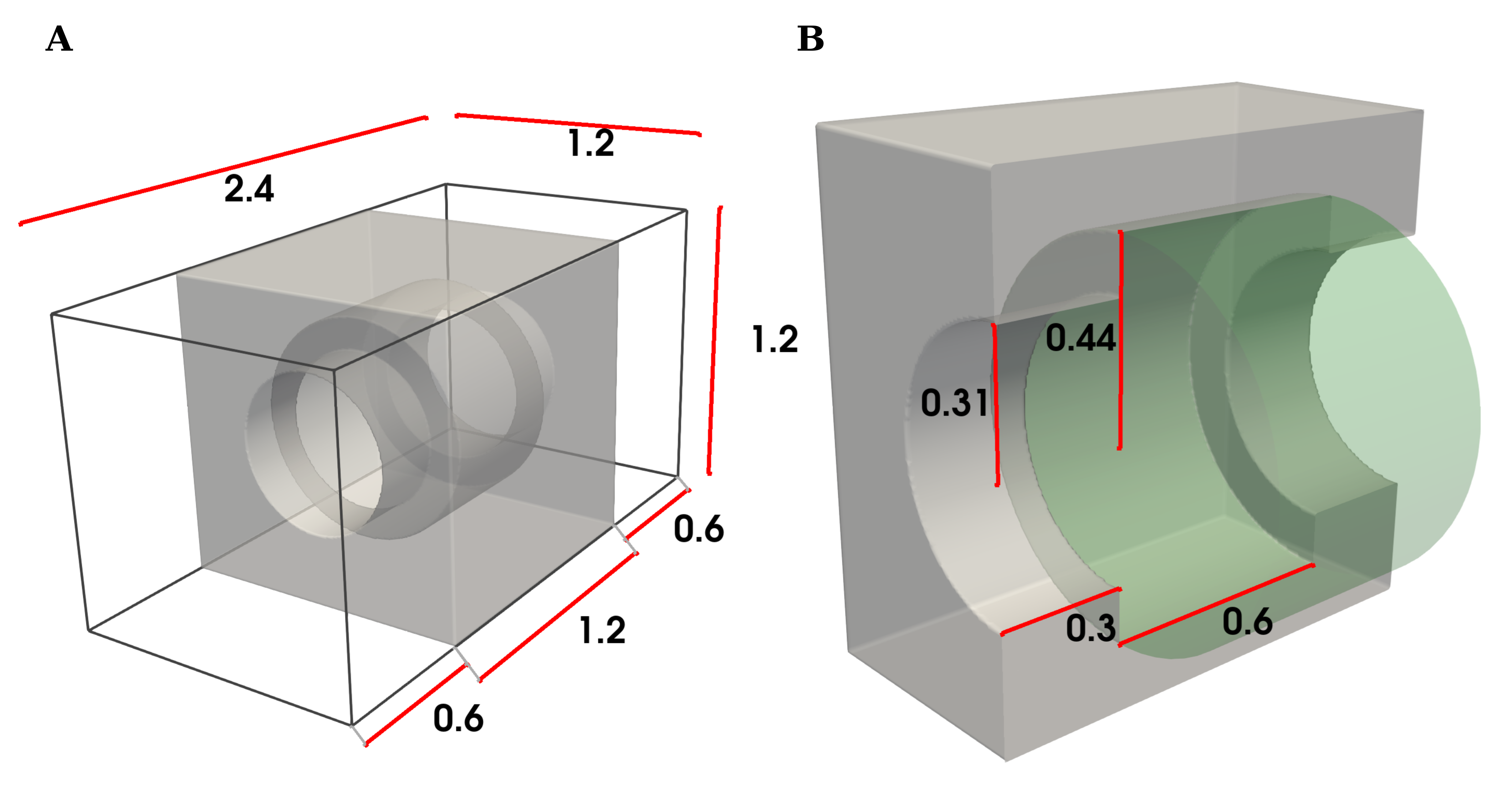}
            \caption{Simulation domain~(A) with the encapsulated monolith (shaded green cylinder) and its support~(B). Length units are centimetres.}
            \label{fig:schema_setup}
        \end{figure}

        This ensures consistency of the imposed boundary conditions at the edges. The specific polynomial functions do not significantly affect the solution as the velocity profile reorganises upstream of the porous medium.
        \textbf{Figure~\ref{fig:schema_setup}A} shows the simulation domain. \textbf{Figure~\ref{fig:schema_setup}B} shows the immersed monolith (in green) and its support, defined by boolean operations (see \textbf{\ref{sec:composite_solid_definition}}).
    
        \begin{table}[htpb!]
            \centering
            \caption{PRCFD parameters for simulation control for the reactive flow through porous media, where length units are centimetres.}
            \label{tab:lethe_simulation_control_parameters}
            \begin{tabular}{c c c}
                \toprule
                \textbf{Parameter} & \textbf{Step 1} & \textbf{Step 2} \\
                \cmidrule(lr){1-3}
                  Time integration scheme & BDF1             & BDF1 \\
                  Element type            & Q$_1$Q$_1$Q$_1$   & Q$_1$ (reactant)  \\ 
                  
                  Initial solution        & $\bm{u}=(u_x,0,0)$ & From Step 1 \\
                                          & $u_x$ from Eq.\ref{eq:imposed_velocity_profile} &  \\
                                          & $c=0$            &            \\
                  Inlet boundary          & $\bm{u}=(u_x,0,0)$ & $c=1$ \\
                                          & $u_x$ from Eq.\ref{eq:imposed_velocity_profile} &  \\
                                          & $c=0$            &         \\
                  Side wall boundary      & $\bm{u}=\bm{0}$ & $\bm{n} \cdot \nabla c= 0$   \\
                                          & $\bm{n} \cdot \nabla c= 0$ &  \\
                  Outlet boundary         & $\nu \nabla \mathbf{u} \cdot \mathbf{n} - p^* \mathcal{I} \cdot \mathbf{n} = \bm{0}$ &  $\bm{n} \cdot \nabla c= 0$ \\
                                          & $\bm{n} \cdot \nabla c= 0$ &   \\
                  Domain dimensions $[\SI{}{\centi\metre}]$ & $[2.4,1.2,1.2]$  & $[2.4,1.2,1.2]$    \\
                \bottomrule
            \end{tabular}
        \end{table}
    
        \begin{table}[htpb!]
            \centering
            \caption{PRCFD physical parameters and dimensionless numbers for the reactive flow through porous media.}
            \label{tab:lethe_physical_properties}
            \begin{tabular}{c c c }
                \toprule
                \textbf{Physical property} & \textbf{Unit} & \textbf{Value} \\
                \cmidrule(lr){1-3}
                  Kinematic viscosity     &  \SI{}{\centi\metre\squared\per\second} & $\nu = 1 \cdot 10^{-2}$  \\ 
                  
                  Diffusivity inside      &  \SI{}{\centi\metre\squared\per\second} & $\mathcal{D}_-  = 1\cdot 10^{-9}$            \\
                  Diffusivity outside     &  \SI{}{\centi\metre\squared\per\second} & $\mathcal{D}_+ = 6.85\cdot 10^{-6}$         \\
                  Diffusivity thickness   &  \SI{}{\centi\metre} & $\varepsilon_\mathcal{D}    = 1\cdot 10^{-3}$            \\
                  
                  Reaction constant at interface  &  \SI{}{\per\second} & $k_0 \in \{1\cdot 10^{-2},\,3\cdot 10^{-2},\,1\cdot 10^{-1},\,3\cdot 10^{-1},\,1\}$         \\
                  Reaction constant thickness  &  \SI{}{\centi\metre} & $\varepsilon_k              \in \{1.25\cdot 10^{-3},\,2.5\cdot 10^{-3},\,5\cdot 10^{-3},\,1\cdot 10^{-2},\,2\cdot 10^{-2}\}$            \\
                  Flow rate  &  \SI{}{\milli\litre\per\second} & $Q \in \{0.013, 0.066, 0.130\}$            \\
                  \cmidrule(lr){1-3}
                  \textbf{Dimensionless number} &  \textbf{Symbol} & \textbf{Range} \\
                    \cmidrule(lr){1-3}
                  Reynolds number     & $\mathrm{Re}$ & $0.25-3.7$  \\ 
                  Péclet number       & $\mathrm{Pe}$ & $300-4000$  \\ 
                  Damköhler number    & $\mathrm{Da}$ & $0.01-0.2$  \\ 
                \bottomrule
            \end{tabular}
        \end{table}

    \subsection{Calibration procedure}
    
\begin{figure}[htpb!]
            \centering
            \includegraphics[width=0.9\textwidth]{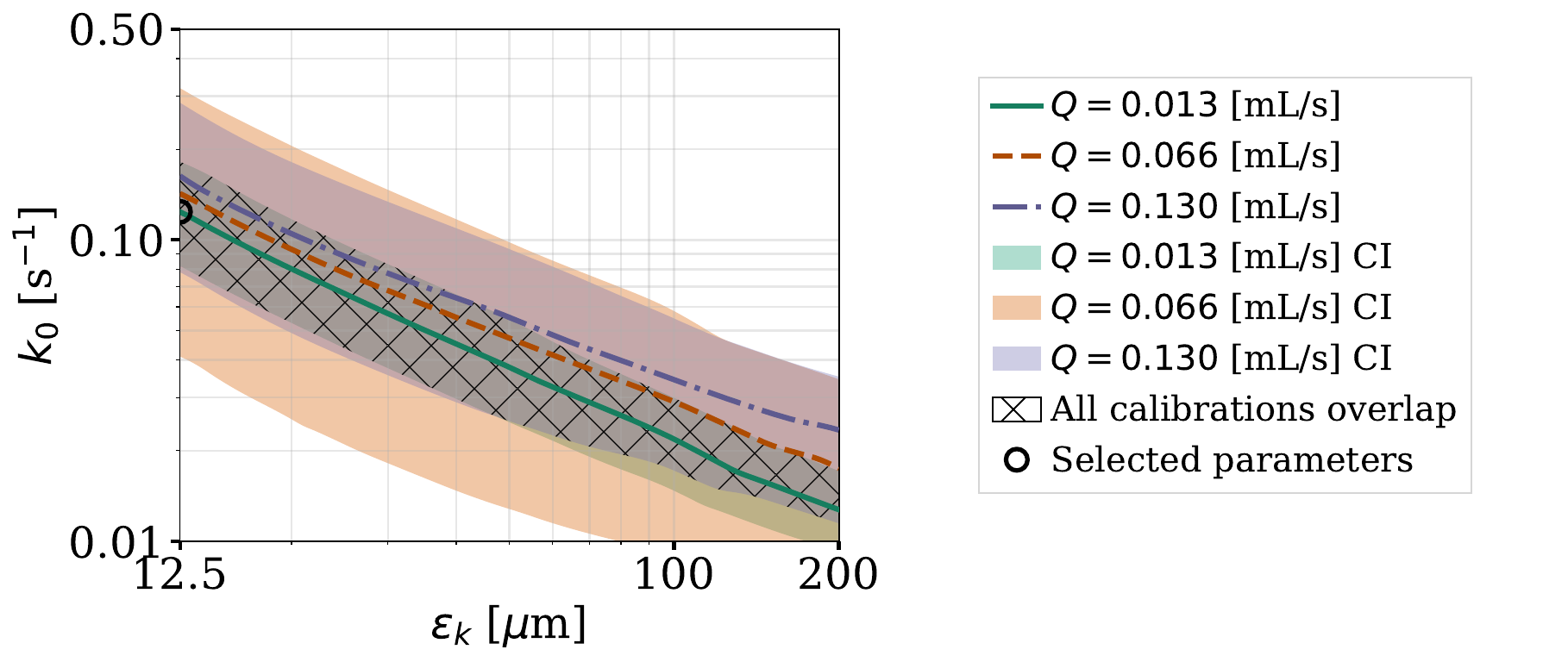}
            \caption{Best fit curves of reaction constants $k_0$ and thicknesses $\varepsilon_k$ where the signed error between experimental and simulated conversions is zero. The curves are interpolated from $5 \times 5$ parametric sweeps and based on the conversion data from the sample $60 \, \mathrm{min}$~\#2. The shaded bands represent the confidence interval~(CI) obtained from the experimental conversions.} 
            \label{fig:combined_calibration_60min2}
        \end{figure}

        We arbitrarily used the experimental conversion-flow rate curves from the sample $60 \, \mathrm{min}$~\#2 to fit the effective reaction constant $k_0$ and reactive shell thickness $\varepsilon_k$, and show the results in \textbf{Figure~\ref{fig:combined_calibration_60min2}}. Each curve corresponds to the best fit for its flow rate $Q$ and is obtained from the interpolation of the signed conversion error (experimental minus simulated), shown in more details in \textbf{\ref{sec:calibration_data_per_flow_rate}}. 
        The shaded areas represent the confidence intervals~(CI) for each curve and are based on the experimental CI. 
    
        We observe that the overlap region is significant and that the curves are closer to each other at lower $\varepsilon_k$ values, where the reaction shell is the thinnest and behaves more like a surface (heterogeneous) reaction. 
        We do not claim that the calibration sets are unique nor that they represent intrinsic kinetics: we identify families of parameters that match experimental data, given how monolith structure is the dominant factor that dictates performance.
        We selected a reaction constant of $k_0 = 0.124$~\SI{}{\per\second} and a thickness $\varepsilon_k = 12.5$~\SI{}{\micro\metre} (shown in \textbf{Figure~\ref{fig:combined_calibration_60min2}}) because it fits all flow rates and corresponds to the minimal $\varepsilon_k$ (most surface-localised reaction) in the sweep.

    \subsection{Transferability validation results}

\begin{figure}[htpb!]
            \centering
            \includegraphics[width=0.7\textwidth]{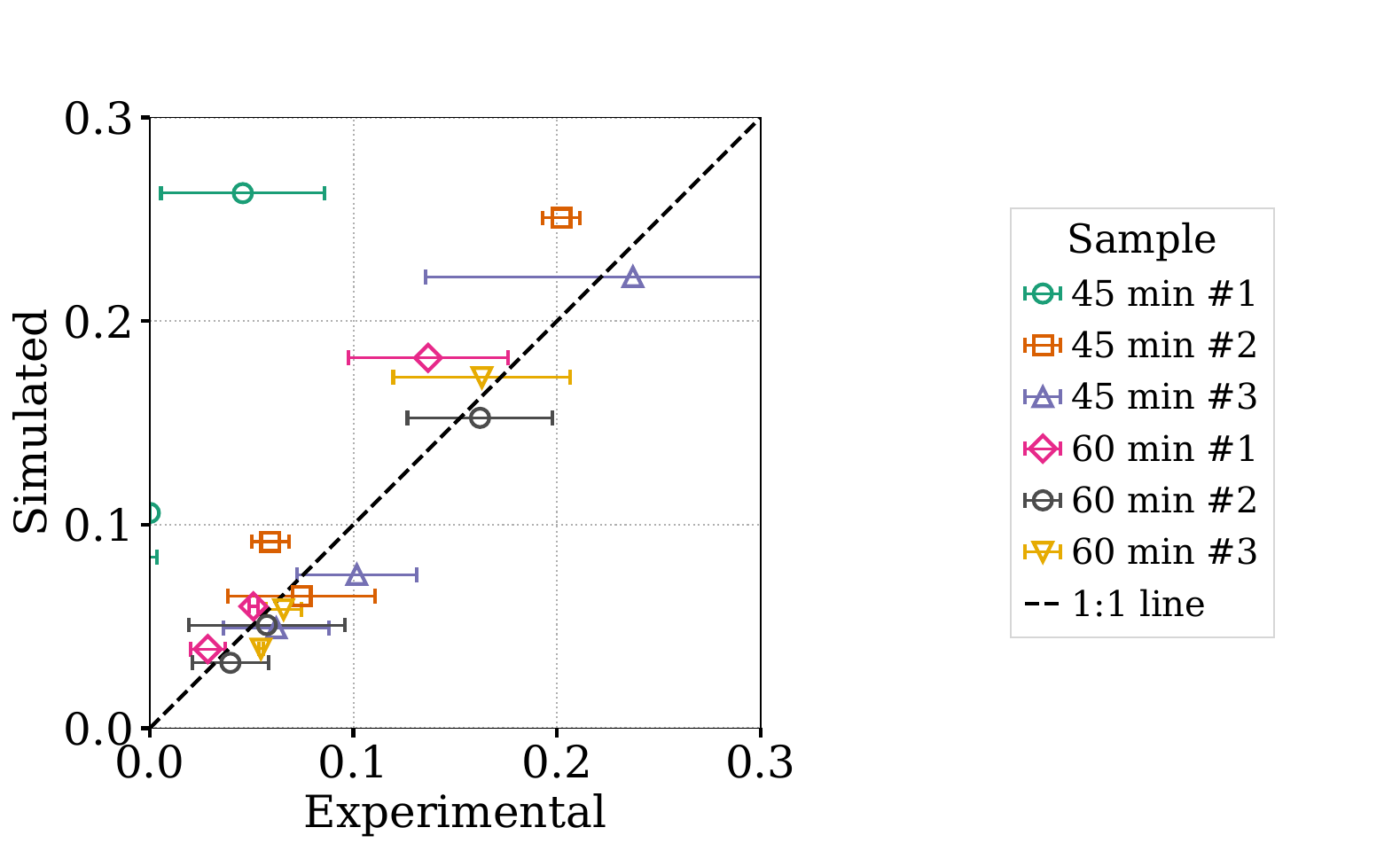}
            \caption{Pareto plot of experimental-simulated pairs of conversions, at flow rates of $Q=\{0.013, 0.066, 0.130\}$~\SI{}{\milli\litre\per\second} for samples $45 \, \mathrm{min}$~\#1$-$3 and $60 \, \mathrm{min}$~\#1$-$3. The model uses a reaction constant of $k_0 = 0.124$~\SI{}{\per\second} and a thickness $\varepsilon_k = 12.5$~\SI{}{\micro\metre}.} 
            \label{fig:pareto_transferability}
        \end{figure}

        We validated the calibration by simulating the reaction at other flow rates and for other samples. \textbf{Figure~\ref{fig:pareto_transferability}} shows the Pareto plot of experimental-simulated pairs of conversions, at flow rates of $Q=\{0.013, 0.066, 0.130\}$~\SI{}{\milli\litre\per\second} for the samples that were not used for calibration, as well as for the sample \SI{60}{\minute}~\#2, shown for reference. 
        
        The model predicts higher conversions than those obtained experimentally for the sample \SI{45}{\minute}~\#1. We attribute this difference to the presence of a polymer film observed after synthesis (see \textbf{\ref{sec:comparison_45min_samples}}). 
        We hypothesise that this contaminant covers the catalytic sites on the PdNPs, thereby reducing their activity.
        A good agreement was obtained between the simulation and the experiments for the remaining data ($5/6$ samples).
        Experimental error likely arises from imperfect bubble removal, channelling effects in the soft silicone, and precision of UV-vis measurements.  Overall, the PRCFD model can predict performance across monoliths and flow rates.

\section{Evidence of surface-access-boundedness}
    \label{sec:diffusion_boundedness}

    The region of acceptable $k_0 - \varepsilon_k$ in \textbf{Figure~\ref{fig:combined_calibration_60min2}} spans the whole range of tested thicknesses. The low impact of the kinetic model parameters on the predictivity of the model hints that the reaction is not \textit{kinetics-limited}. To verify this hypothesis, we performed a batch experiment using sample \SI{60}{\minute}~\#3 in a glass vial for a duration of \SI{60}{\minute}. The results are detailed in \textbf{\ref{sec:batch_reaction_results}}.
    The conversion reaches a plateau at $0.7$, indicating that the conversions reported in \textbf{Figure~\ref{fig:experimental_conversion}} are well below the kinetic limit and are therefore not controlled by intrinsic reaction kinetics. 
    The plateau being lower than $1$ is likely due to the partial depletion of NaBH$_4$ or parasitic reactions, which would affect the equilibrium and invalidate kinetics simplifications. However, this does not affect the validity of the present analysis, since the conversions in \textbf{Figure~\ref{fig:experimental_conversion}} remain below $0.25$, where the above assumptions hold.
    
    We verified that the reaction setup is not \textit{diffusion-limited} using three strategies:
    \begin{enumerate}
        \item Multiplication of the diffusivity by factors of $\{1,2,4,8\}$ to verify the magnitude of its impact on overall conversion;
        \item Reduction of the transition thickness  $\varepsilon_{\mathcal{D}}$ to 10\% of its initial value in the diffusivity transition model (Equation~\ref{eq:sdf_dependent_diffusivity}) to ensure that the diffusivity smoothing across phases is not excessive,  which would otherwise artificially lower diffusivity near the surface and hinder overall conversion;
        \item Removal of the diffusivity transition model (Equation~\ref{eq:sdf_dependent_diffusivity}) to exclude it as a potential source of error ($\mathcal{D}_\mathrm{-} = \mathcal{D}_\mathrm{+}$).
    \end{enumerate}
    We simulated the reactive flow through sample \SI{60}{\minute}~\#2 at $Q=$~\SI{0.013}{\milli\litre\per\second} while applying the three strategies, and show the results in \textbf{Table~\ref{tab:diffusivity_toy_case}}.

    \begin{table}[htpb!]
        \centering
        \caption{Effect of diffusivity on conversion for the low-flow reactive PRCFD case.}
        \label{tab:diffusivity_toy_case}
        \begin{tabular}{c c c c}
            \toprule
            \multirow{1}{*}{\bfseries Diffusivity multiplier} 
              & \multicolumn{3}{c}{\textbf{Conversion} } \\
            \cmidrule(lr){2-4}
              \textbf{ } & \textbf{Base} & \textbf{Thin transition} & \textbf{No transition} \\
              \textbf{ } & \textbf{ } & $0.1\varepsilon_{\mathcal{D}}$ &  $\mathcal{D}_\mathrm{-} = \mathcal{D}_\mathrm{+}$ \\
            \hline
              $1$ & 0.1599 & 0.1545 & 0.1708 \\
              $2$ & 0.1749 & 0.1667 & 0.1910 \\
              $4$ & 0.1908 & 0.1794 & 0.2128 \\
              $8$ & 0.2066 & 0.1919 & 0.2350 \\
            \bottomrule
        \end{tabular}
    \end{table}
    
    The conversion increases modestly with an increase in molecular diffusivity. Using a thinner transition zone decreases slightly the conversion, but its effect is minor. Removing the transition entirely increases slightly the conversion, but this small increase confirms that molecular diffusivity is not the limiting factor to the conversion. 
    
    After exclusion of both kinetics-limited and diffusion-limited behaviours, we made the hypothesis that advective mixing and flow distribution play a dominant role in controlling conversion. We tested this hypothesis by comparing the effects of the Reynolds and Péclet numbers, taken here as analogues of advection and diffusion, respectively. We compared the conversion of sample \SI{60}{\minute}~\#2 for three cases: $(\mathrm{Re},\mathrm{Pe})=(0.36,530)$, $(3.6,530)$, and $(3.6,5300)$, obtained by varying the flow rate and diffusivity. 
    We show the results in \textbf{Table~\ref{tab:conversion_constant_pe}}. 
    Cases (1) and (2), which have the same Péclet number but different Reynolds numbers, show markedly different conversions. By contrast, cases (2) and (3), which have the same Reynolds number but Péclet numbers differing by a factor of 10, show similar conversions. This indicates that, in this regime, conversion is more sensitive to the Reynolds number than to the Péclet number. We conclude that the flow rate, and by extension the flow patterns, have the strongest influence on the conversion.
    
    \begin{table}[htpb!]
        \centering
        \caption{Conversion through sample \SI{60}{\minute}~\#2 at varying flow rates and diffusivities. The Péclet number is matched between (1) and (2) by adjusting flow rate and diffusivity. The Reynolds number is matched between (2) and (3).}
        \label{tab:conversion_constant_pe}
        \begin{tabular}{l c c c c c}
            \toprule
            & \textbf{Re} & \textbf{Pe} & \textbf{Flow rate} & \textbf{Diffusivity} & \textbf{Conversion} \\
            &  &   & [\SI{}{\milli\litre\per\second}] & [\SI{}{\micro\metre\squared\per\second}] &   \\
            \cmidrule(lr){2-6}
              (1) Base case & $0.36$ & $530$ & $0.013$  & $685$  & $0.1749$ \\
              (2) Increased $\mathrm{Re}$, constant $\mathrm{Pe}$ & $3.6$ &  $530$ & $0.130$  & $6850$  & $0.0280$ \\
              (3) Increased $\mathrm{Re}$, increased $\mathrm{Pe}$ & $3.6$ & $5300$ & $0.130$  & $685$  & $0.0317$ \\
            \bottomrule
        \end{tabular}
    \end{table}

\section{Performance comparison across geometries}
    \label{sec:structural_design}

    \subsection{Geometries and similar conditions}

        Evidence reveals how geometry-induced surface access controls the efficiency of \textit{p}-nitrophenol reduction catalysed by PdNP@silicone monoliths. 
        To highlight its importance, we generated alternative structures using the Triply Periodic Minimal Surfaces~(TPMS) generator Cesogen~\cite{patience2024cesogen} and a packed bed using the discrete element method implemented in Lethe~\cite{alphonius2025lethe}. We then resolved the reactive flow using the same hydrodynamics conditions and kinetic parameters.
        
        We generated the structures to display matched macroscopic properties as the synthesised \SI{60}{\minute} samples: porosity of 50\%, surface of \SI{8}{\centi\metre\squared}, cylindrical shape of diameter \SI{0.88}{\centi\metre} and height \SI{0.6}{\centi\metre}. In addition to these structured media, we generated structures with doubled surface (\SI{16}{\centi\metre\squared}) to compare more fairly to samples with annealing durations of \SI{45}{\minute}.
        The packed bed has a lower porosity~(33\%) due to the geometric constraints of spherical packings.
        The TPMS-based structures are built from the intersection of a box and TPMS: gyroid, multiscale-gyroid, Schwarz-diamond, Schwarz-P, IWP. \textbf{Table~\ref{tab:samples_summary}} describes macroscopic properties of each geometry. The generated structured media are shown in \textbf{\ref{sec:generated_structured_media}}.

        \begin{table}[htpb!]
            \centering
            \caption{Summary of porous media samples and associated properties. TPMS endoskeletons are defined with $C_{\{x,y,z\}}:=\cos (2\pi\{x,y,z\})$ and $S_{\{x,y,z\}}:=\sin (2\pi\{x,y,z\})$, then scaled and thickened/eroded to matched target porosity and surface.}
            \label{tab:samples_summary}
            \begin{tabular}{l l c c}
                \toprule
                \textbf{Name} & \textbf{Generation} & \textbf{Surface} & \textbf{Porosity}\\
                  &   & [\SI{}{\centi\metre\squared}] & [\SI{}{\percent}]\\
                \midrule
                \SI{45}{\minute}~\#1 & Polymers mix annealed for \SI{45}{\minute} & 23 & 45 \\
                \SI{45}{\minute}~\#2 & Polymers mix annealed for \SI{45}{\minute} & 17 & 40 \\
                \SI{45}{\minute}~\#3 & Polymers mix annealed for \SI{45}{\minute} & 10 & 45 \\
                \SI{60}{\minute}~\#1 & Polymers mix annealed for \SI{60}{\minute} & 8 & 53 \\
                \SI{60}{\minute}~\#2 & Polymers mix annealed for \SI{60}{\minute} & 8 & 49 \\
                \SI{60}{\minute}~\#3 & Polymers mix annealed for \SI{60}{\minute} & 7 & 46 \\
                Simple, type D & TPMS: $S_xS_yS_z+C_xS_yC_z+C_yS_zC_x+C_zS_xC_y$& 8 & 50 \\
                Simple, type Gyroid & TPMS: $C_xS_y+C_yS_z+C_zS_x$  & 8 & 50 \\
                Simple, type Multiscale Gyroid & Intersection of two scales of gyroid: $1$ and $0.5$ & 8 & 50 \\
                Simple, type IWP & TPMS: $2(C_xC_y+C_yC_z+C_zC_x)-(C_{2x}+C_{2y}+C_{2z})$ & 8 & 49 \\
                Simple, type P & TPMS: $C_x+C_y+C_z$ & 8 & 50 \\
                Double, type D & TPMS: $S_xS_yS_z+C_xS_yC_z+C_yS_zC_x+C_zS_xC_y$& 16 & 49 \\
                Double, type Gyroid & TPMS: $C_xS_y+C_yS_z+C_zS_x$  & 16 & 48 \\
                Double, type Multiscale Gyroid & Intersection of two scales of gyroid: $1$ and $0.5$ & 16 & 49 \\
                Double, type IWP & TPMS: $2(C_xC_y+C_yC_z+C_zC_x)-(C_{2x}+C_{2y}+C_{2z})$ & 16 & 48 \\
                Double, type P & TPMS: $C_x+C_y+C_z$ & 16 & 48 \\
                Spheres of diameter \SI{1.2}{\milli\metre} & Discrete element method & 13 & 33 \\
                %Porosité parcked bed 32.734% selon paraview
                \bottomrule
            \end{tabular}
        \end{table}

    \subsection{Conversion and productivity}

\begin{figure}[htpb!]
            \centering
                \includegraphics[width=1.0\textwidth]{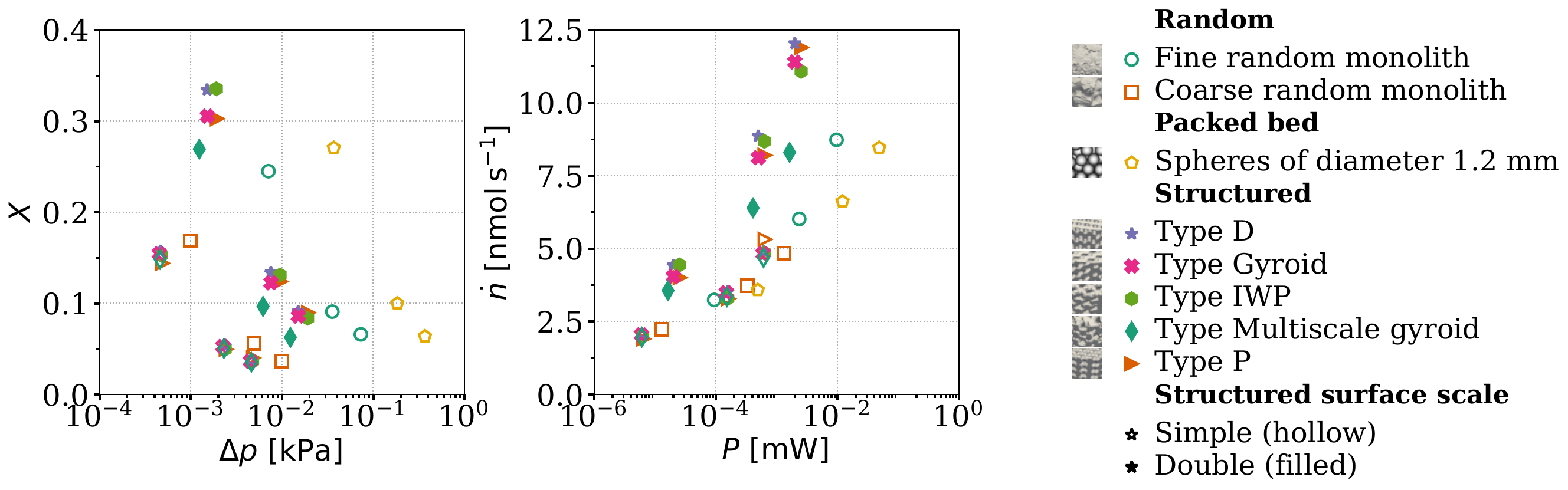}
            \caption{Performance of digitised random monoliths, TPMS-based structures (simple and double surface) and a packed bed as monolith reactors predicted by PRCFD. From left to right: reactor conversion versus pressure drop, molar production rate versus required pumping power.}
            \label{fig:productivity_vs_pumping_power}
        \end{figure}

        \textbf{Figure~\ref{fig:productivity_vs_pumping_power}} shows the performance of each examined structure as a monolith reactor by comparing the conversion~$X$ as a function of the pressure drop~$\Delta p$, as well as the molar production rate~$\dot{n}$ as a function of the required pumping power~$P$. The molar production rate is computed as:
    
        \begin{align}
             \dot{n}= X Q c_\mathrm{0} , 
            \label{eq:productivity}
        \end{align}
        based on the conversion $X$, volumetric flow rate $Q$, and inlet concentration $c_\mathrm{0}$.
        The pumping power is computed as
        
        \begin{align}
            P= Q \Delta p.
            \label{eq:pumping_cost}
        \end{align}
        
        The relationship between $X$ and $\Delta p$ is similar within each surface-scale group: coarse monoliths, comprising simple structured media and coarse random monoliths (\SI{60}{\minute} samples), and fine monoliths, comprising double structured media and fine random monoliths (\SI{45}{\minute} samples). 
        Comparisons are therefore made between structures with similar total surface area: simple TPMS against the \SI{60}{\minute} random monoliths, and double-surface TPMS against the higher-surface \SI{45}{\minute} random monoliths.
        At comparable $X$ values, the $\Delta p$ of random monoliths is higher by a factor of $5-10$. The packed bed shows $X$ values similar to those of the fine monoliths, but at much higher $\Delta p$, which is explained by its lower porosity (33\%).
        The $\dot{n}-P$ relations of the coarse monoliths are also similar despite their different internal geometries. In contrast, the fine structured monoliths clearly outperform the fine random monoliths: for an equivalent $\dot{n}$ value (e.g., $\dot{n}=\SI{9}{\milli\mol\per\second}$), the required $P$ differs by more than one order of magnitude between the TPMS-based structures and the fine random monoliths. $\dot{n}$ also increases more rapidly with $P$ for the fine structured media than for the other media. Finally, at constant $P$, the fine structured media achieve distinct $\dot{n}$ values, whereas the coarse media tend to collapse onto similar performance. 
        These results suggest that reactor performance becomes more sensitive to internal geometry as surface area increases and pore size decreases, because smaller flow passages increase hydraulic resistance and amplify flow maldistribution, making access to the catalytic surface more uneven.

    \subsection{Reaction efficiency and surface access}

\begin{figure}[htpb!]
            \centering
            \includegraphics[width=\textwidth]{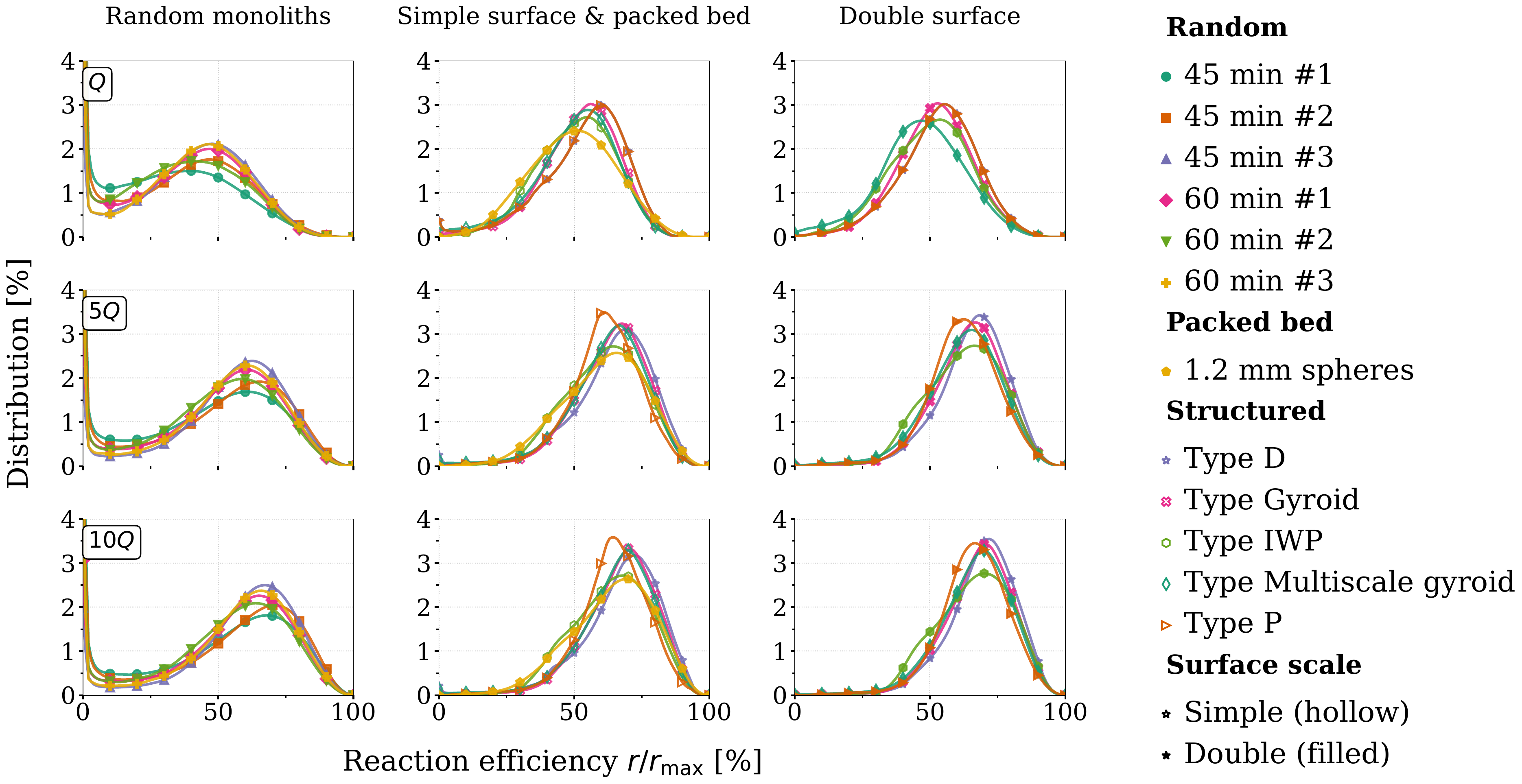}
            \caption{Surface distribution of the reaction efficiency across digitised random monoliths (\SI{45}{\minute}~\#$1-3$, \SI{60}{\minute}~\#$1-3$), structured TPMS-based monoliths (simple and double surface types D, Gyroid, IWP, Multiscale gyroid, P), and a packed bed.
            Each row corresponds to a flow rate, where the base flow rate $Q=0.013$~\SI{}{\milli\litre\per\second}. 
            The first column displays the random monoliths, the second the simple surface TPMS-based structures and the packed bed, and the third the double surface TPMS-based structures.
            }
            \label{fig:surface_efficiency_cdf}
        \end{figure}

        We used Paraview~\cite{paraview2015} to compute the reaction rate across the entire surface of each sample and normalise the computed reaction rate by its theoretical maximum computed as $r_\mathrm{max}=k_0c_0$, based on Eq.~\ref{eq:reaction_rate}.
        \textbf{Figure~\ref{fig:surface_efficiency_cdf}} shows the (surface) probability density function based on local reaction efficiency. We show the cumulative distribution function in \textbf{\ref{sec:surface_distribution_function}}. 
        
        Pore-scale analysis of the reaction rate efficiency $r/r_\mathrm{max}$ confirms that the utilisation of the surface varies depending on the structure and on the flow rate. 
        Each random digitised monolith's distribution shows that a significant fraction of the surface operates below 10\% efficiency, then shows a wide peak between $50\%-70\%$. In contrast, the TPMS-based structures and the packed bed exhibit much lower fractions of underused surface and each a single peak after 50\%, which is both higher and narrower than those of the random monoliths. 
        \textbf{Table~\ref{tab:scalar_surface_utilisation}} supports this interpretation: for each porous medium and flow rate, we show the fraction of the catalytic surface which has a reaction efficiency above 50\%. 
        Generally, increasing flow rate increases the surface efficiency, possibly due to the flow being more constrained to use otherwise avoided paths. 
        Within each family of structured media, the higher-surface versions are more sensitive to the flow rate: at low flow rates, the finer structures show lower surface utilisation, but this behaviour is reversed at higher flow rates. 
        Increasing the surface area within a constant reactor volume does not proportionally increase the molar production rate, because part of the additional surface may remain poorly accessed under low-flow conditions.
        Overall, the structured media exhibit a higher surface access, although the difference is more pronounced at lower flow rates. 
        
        \begin{table}[htpb!]
            \centering
            \caption{Percentage of each porous medium's surface where the reaction rate is higher than 50\% of the maximal reaction rate, given the inlet concentration and kinetics parameters. Each column corresponds to a different flow rate, based on $Q=0.013$~\SI{}{\milli\litre\per\second}.}
            \label{tab:scalar_surface_utilisation}
            \begin{tabular}{l c c c}
                \toprule
                \textbf{Porous medium} & \textbf{$Q$} & \textbf{$5Q$} & \textbf{$10Q$}\\
                 & [\SI{}{\percent}] & [\SI{}{\percent}] & [\SI{}{\percent}]\\
                \midrule
                \SI{45}{\minute}~\#1 & 24 & 51 & 60 \\
                \SI{45}{\minute}~\#2 & 32 & 59 & 66 \\
                \SI{45}{\minute}~\#3 & 38 & 68 & 76 \\
                \SI{60}{\minute}~\#1 & 34 & 60 & 68 \\
                \SI{60}{\minute}~\#2 & 29 & 55 & 64 \\
                \SI{60}{\minute}~\#3 & 36 & 63 & 71 \\
                Spheres of diameter \SI{1.2}{\milli\metre} & 50 & 76 & 81 \\
                Type D (simple) & 65 & 84 & 88 \\
                Type D (double) & 62 & 89 & 93 \\
                Type Gyroid (simple) & 60 & 85 & 90 \\
                Type Gyroid (double) & 56 & 87 & 92 \\
                Type IWP (simple) & 54 & 77 & 81 \\
                Type IWP (double) & 52 & 80 & 86 \\
                Type Multiscale gyroid (simple) & 56 & 83 & 88 \\
                Type Multiscale gyroid (double) & 43 & 83 & 89 \\
                Type P (simple) & 65 & 83 & 88 \\
                Type P (double) & 62 & 86 & 91 \\
                \bottomrule
            \end{tabular}
        \end{table}

\begin{figure}[htpb!]
            \centering
            \includegraphics[width=0.8\textwidth]{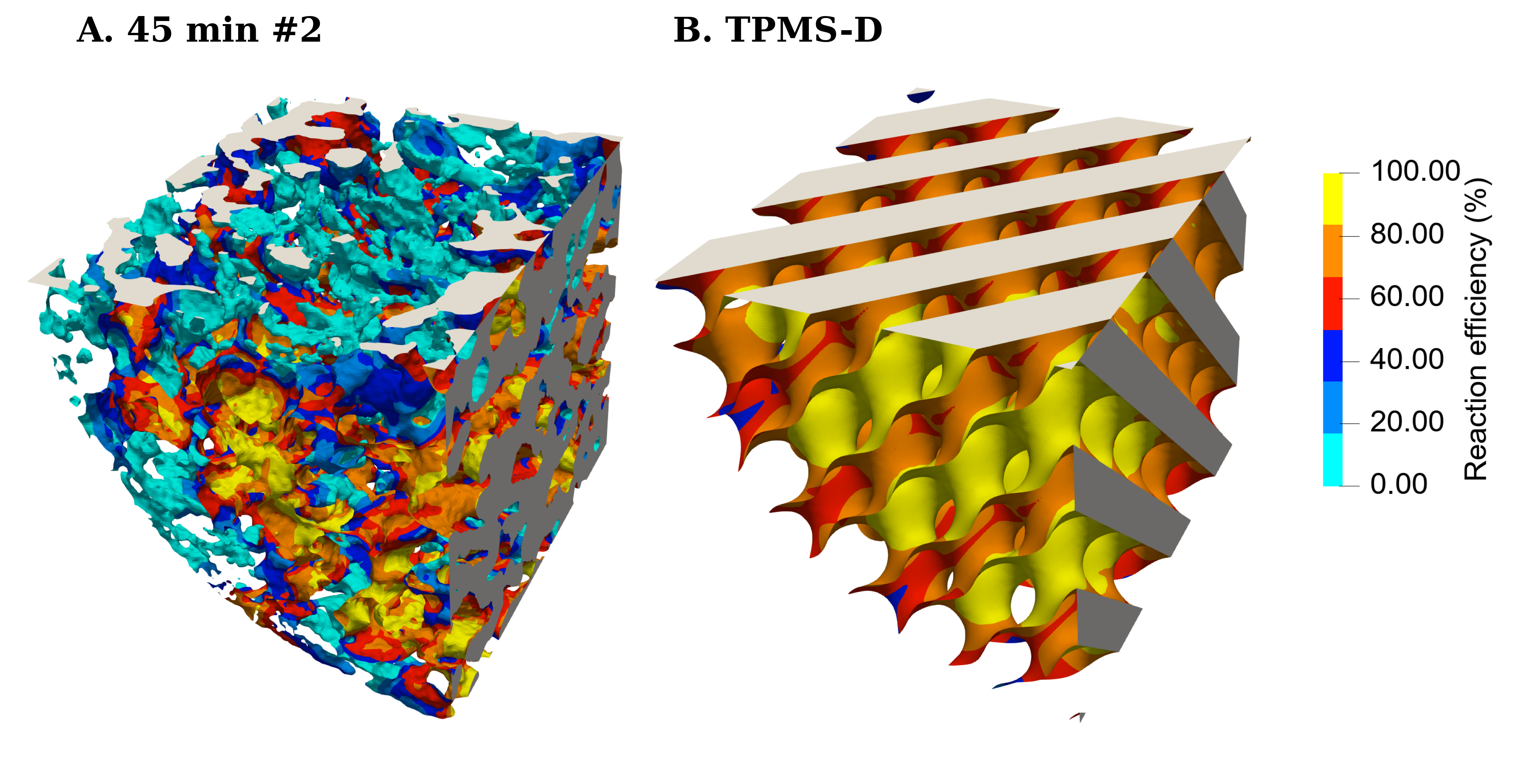}
            \caption{Local reaction efficiency of the sample \SI{45}{\minute}~\#2~(A) and the simple TPMS-based~Type~D structure~(B), at a flow rate of \SI{0.013}{\milli\litre\per\second}. Each structure corresponds to a quarter of its source geometry.
            } \label{fig:quarter_view_45min2_tpmsd_0_013mls_rate}
        \end{figure}

        Visualisation of the reaction rate at the surface clarifies the difference in surface usage; \textbf{Figure~\ref{fig:quarter_view_45min2_tpmsd_0_013mls_rate}} shows the difference in local reaction rates between the sample \SI{45}{\minute}~\#2 and the simple TPMS-based~Type~D structure, at the minimum flow rate of $Q=0.013$~\SI{}{\milli\litre\per\second}. The sample \SI{45}{\minute}~\#2~(A) exhibits wide variations in reaction rates, with large regions below 20\%~efficiency. In contrast, the TPMS-based~Type~D shows more uniform reaction rates.

\begin{figure}[htpb!]
            \centering
            \includegraphics[width=0.7\textwidth]{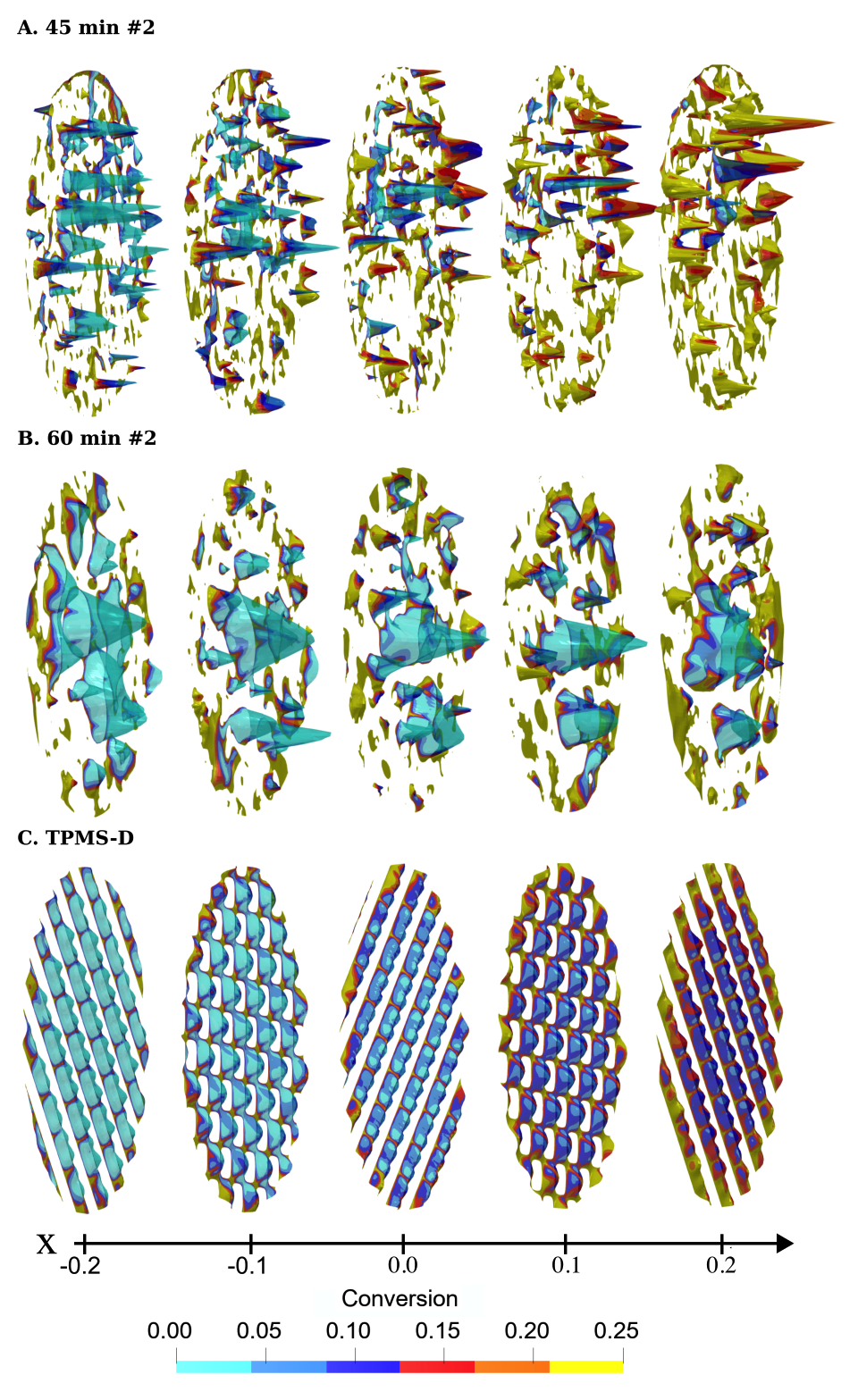}
            \caption{Slices across the flow axis X of \SI{45}{\minute}~\#2~(A), \SI{60}{\minute}~\#2~(B), and the simple TPMS-based~Type~D structure~(C), at a flow rate of \SI{0.013}{\milli\litre\per\second}. 
            Each slice shows the fluid phase coloured by conversion and is deformed according to the local Reynolds~number~$\mathrm{Re}=\lambda u_x / \nu$, where $\lambda$ is the signed distance, $u_x$ is the velocity component along the flow axis, and $\nu$ is the kinematic viscosity. The slices are taken at $x=\{-0.2,-0.1,0,0.1,0.2\}$ \SI{}{\centi\metre}.
            } \label{fig:slices_45min2_60min2_tpmsd_0_013mls_rate}
        \end{figure}

        In line with \textbf{Figure~\ref{fig:quarter_view_45min2_tpmsd_0_013mls_rate}}, \textbf{Figure~\ref{fig:slices_45min2_60min2_tpmsd_0_013mls_rate}} shows sliced views along the flow axis in the geometries \SI{45}{\minute}~\#2, \SI{60}{\minute}~\#2, and TPMS-based~Type~D. 
        The slices are coloured by the local conversion and deformed according to the local Reynolds~number~$(=\lambda u_x / \nu)$, allowing identification of regions that combine high conversion with efficient downstream transport.
        \SI{45}{\minute}~\#2~(A) shows high Reynolds peaks with high conversion when compared to other geometries but, also importantly, wide Reynolds variations that can be associated with high pumping costs~\cite{guevremont2024poreresolvedcfddigitaltwin}. 
        In contrast, \SI{60}{\minute}~\#2~(B) shows clear channelling through its large low-conversion pores, allowing fluid to pass rapidly through weakly reactive regions and transport unreacted species downstream, thereby reducing the overall conversion.
        TPMS-based~Type~D~structure~(C) shows the most uniform profiles across each slice, with steadily progressing reaction; the absence of any outlying peak is consistent with the periodic structure and allows conversion comparable to the \SI{60}{\minute}~samples, at reduced pumping costs. 
        
        The flow maldistribution in the digitised random samples results in worse performance overall when compared to the structured media. The fine random monoliths, although they have high molar production rate thanks to their high specific surface, have high pumping costs due to their small pore size, but also due to channelling. The coarse monoliths' channelling increases pumping costs while limiting access to part of their catalytic surface: they therefore have lower molar production rate and higher pumping costs than the structured monoliths. Interestingly, the random monoliths' molar production rate-pumping cost curves appear to collapse onto one another in Figure \ref{fig:productivity_vs_pumping_power}, indicating that increased flow rate through the coarse random samples may result in similar productivity. This tendency may be explained by the self-similarity of these media, which have been fabricated using similar protocols with the only difference being the annealing duration.
        The packed bed has high molar production rate but requires by far the most pumping power.

\section{Damköhler number considerations in reactive monoliths}\label{sec:damkohler_considerations}

\begin{figure}[htpb!]
        \centering
            \includegraphics[width=0.8\textwidth]{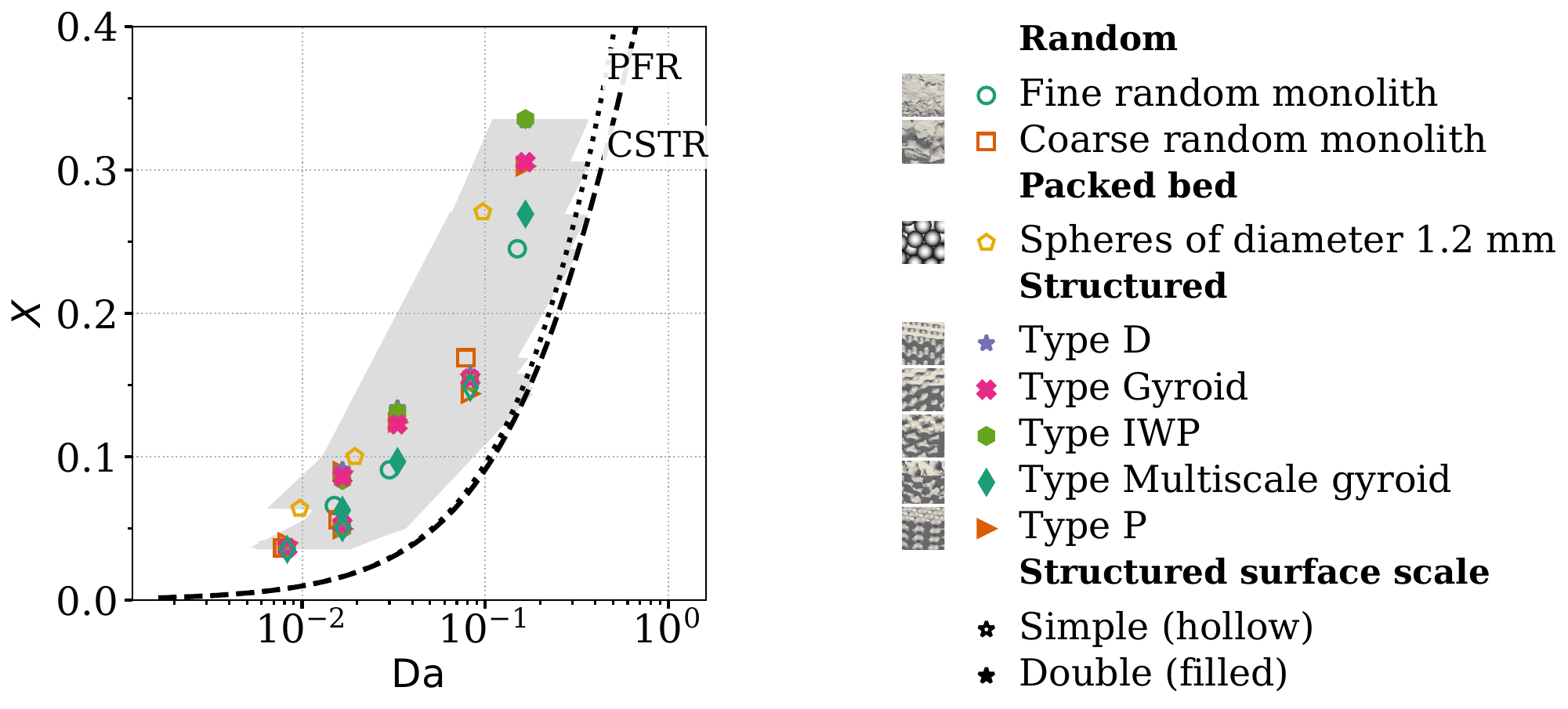}
        \caption{Conversion versus Damköhler number for random monoliths, structured monoliths, and a packed bed. The CSTR and PFR models are reference relations under homogeneous mixing assumptions. The shaded region represent the range covered by the maximum and minimum Damköhler number that can be computed based on the valid calibration region in \textbf{Figure~\ref{fig:combined_calibration_60min2}}.}
        \label{fig:conversion_vs_damkohler}
    \end{figure}

    We showed that, in highly heterogeneous media with strong concentration gradients, predicting reactor performance requires information on surface replenishment and flow uniformity, which is not encompassed by classical regime classification.
    We computed the equivalent Damköhler numbers for each geometry-flow rate pair and summarised the conversion-Damköhler relations in \textbf{Figure~\ref{fig:conversion_vs_damkohler}}, along with the expected conversions for idealised reactors: Continuous Stirred Tank Reactor~(CSTR) and Plug Flow Reactor~(PFR). For a given conversion, the matching Damköhler number depends on the chosen calibration pair $k_0 - \varepsilon_k$ (see Eq.~\ref{eq:damkohler}); 
    the shaded region reflects the impact of this choice taken from the overlap region in \textbf{Figure~\ref{fig:combined_calibration_60min2}}.

    Both CSTR and PFR models are based on homogeneity assumptions: CSTR is perfectly mixed through its whole volume, and PFR has no radial gradients, only axial. 
    The conversions predicted through PRCFD for all the media lie above the CSTR and PFR relations for the same reported Damköhler numbers.
    We explain this mismatch by the classical assumptions that are not satisfied in our reactors: uniform residence time, absence of flow maldistribution, and spatially homogeneous reaction rate. 
    In contrast, we use a definition of the Damköhler number that collapses a surface-localised reaction into a volumetric averaged process using the bulk residence time and an equivalent volumetric rate from eggshell parameters; it does not capture how localised reaction zones experience concentrations that differ from the bulk-averaged value.
    This behaviour can be explained by the transport of fresh reactant deeper in the media (through larger channels), the high activity of the surface (when not smoothed out through the volume), and the emergence of high flow/high concentration channels.
    These results may be counterintuitive but are not incompatible with the suboptimal surface accessibility studied throughout the previous sections: conversion is dominated by a few well-fed regions and a small fraction of surface contributes disproportionately.
    The more classical descriptions of reactors based on the Damköhler or Thiele numbers, which compare reaction and transport timescales, become insufficient to fully characterise performance when surface is not uniformly accessible or when maldistribution patterns appear at the reactor scale.

\section{Conclusion}\label{sec:conclusion}

    We fabricated porous monolith supports and synthesised catalytic PdNPs \textit{in situ}, then combined experiments with PRCFD to show how surface-access-boundedness becomes a limiting factor in the performance of porous monolith reactors under advective regimes, especially in unstructured media with wide pore size distributions.
    The structured media we studied had an order of magnitude lower power consumption than randomised media, at equal productivity; this difference can be explained by flow maldistribution in randomised media.
    PRCFD allows prediction of this behaviour and its extent after reaction model calibration with minimal experimental data, whereas models based only on macroscopic descriptors may lack predictability by missing the interplay between flow patterns and structure.
    Geometry governs surface utilisation by shaping flow patterns, such as channelling in some regions and stagnation in others.

    Beyond theoretical considerations, PRCFD supports the design and diagnosis of catalyst support geometry by distinguishing not only between candidate geometries, but also between fabrication methods. 
    The fabrication of structured media at such micrometric scales remains a practical challenge, whereas random porous media are comparatively easier to manufacture.
    Although we have a good understanding of the relationship between the parameters of polymer-mixing-based monoliths and their influence on porosity and specific surface, PRCFD reveals the extent to which wide pore size distributions simultaneously increase pumping costs and reduce surface accessibility compared to alternative structured media. 
    Monolith reactors have high productivity potential, but managing their performance at high flow rates, when flow maldistribution can become more limiting than diffusion or kinetics, requires the level of detail that PRCFD can provide.

    We analyse in depth the impact of catalyst support geometry on the reduction of a single species. The kinetics model is simple and we do not consider multiple species, reactions, nor thermal effects. We assume that the active phase on the support is uniform and that no deactivation or fouling occurs. 
    Recent work in the field includes \textmu CT-based PRCFD with multi-species combustion, without surface catalytic effects~\cite{boigne2024iRmicroCT}, and \textmu CT-based PRCFD CO-oxidation with platinum-NP/$\alpha$-alumina with full thermal effects of a single ceramic open-cell foam monolith~\cite{dong2018rxPRCFD}.
    While our approach cannot distinguish between elementary reactions to identify the kinetic rate-limiting step, it allows fair comparison of multiple geometries and insight at the core of surface-access-boundedness. 

    This work positions porous media geometry as a design variable for process intensification for high-efficiency surface-based operations including filtration, separation and chemicals production. 
    Building upon its role in performance during the operation of porous monolith reactors, future work could integrate PRCFD with geometry-generators and surrogate models to enable optimization under multi-objective constraints. 
    Alternative avenues might include more complex reaction systems to distinguish how structure affects rate-limiting steps and overall selectivity.

\section{CRediT authorship contribution statement}
    \label{sec:credit}
    %Conceptualisation; Data curation; Formal analysis; Funding acquisition; Investigation; Methodology; Project administration; Resources; Software; Supervision; Validation; Visualisation; Writing - original draft; Writing - review \& editing.
    \textbf{Olivier Guévremont}: 
    Conceptualisation, Data curation, Formal analysis, 
    Investigation, Methodology, Software, Validation, Visualisation, 
    Writing - original draft.
    \textbf{Olivier Gazil}: 
    Conceptualisation, Formal analysis,
    Methodology, Validation, Visualisation,
    Writing - review \& editing.
    \textbf{Federico Galli}: 
    Methodology, Resources, Supervision, 
    Writing - review \& editing.
    \textbf{Nick Virgilio}: 
    Funding acquisition, Methodology, Resources, Supervision, 
    Writing - review \& editing.
    \textbf{Bruno Blais}: 
    Conceptualisation, Funding acquisition, Methodology, Resources,
    Software, Supervision, 
    Writing - review \& editing.

\section{Declaration of competing interest}
    \label{sec:declaration_of_interest}
    
    The authors declare the following financial interests/personal relationships which may be considered as potential competing interests:
    Bruno Blais reports financial support was provided by Natural Sciences and Engineering Research Council of Canada through the RGPIN2020-04510 Discovery Grant and the MMIAOW Canada Research Chair (level 2) in Computer-Assisted Design and Scale-up of Alternative Energy Vectors for Sustainable Chemical Processes. 
    Olivier Guévremont reports financial support was provided by Natural Sciences and Engineering Research Council of Canada, Fonds de recherche du Québec - Nature and technologies and Institut de l'Énergie Trottier.
    Bruno Blais and Federico Galli report compute time was provided by Digital Research Alliance of Canada.

\section{Data availability}
    \label{sec:data_availability}
    The data supporting this study are available in a Zenodo repository (DOI: 10.5281/zenodo.19323368). These include imaging datasets, simulation geometries (STL and RBF formats), parameter files, summary results, and experimental data.
    
\section{Acknowledgments}
    \label{sec:acknowledgments}
    
    Olivier Guévremont acknowledges the financial support of the Natural Sciences and Engineering Research Council of Canada (NSERC), the Fonds de recherche du Québec - Nature et technologies (FRQNT), and the Institut de l’Énergie Trottier (IET).
    The authors acknowledge the Digital Research Alliance of Canada for technical support and computing time; the Centre de Caractérisation Microscopique des Matériaux (CM2) for its support with TEM characterisation; and Marie-Hélène Bernier (GCMLab, École Polytechnique de Montréal), as well as Matthieu Gauthier, Sylvain Simard Fleury, and Alejandro Vélez Garcia from the Department of Chemical Engineering, École Polytechnique de Montréal, for their technical support and contributions to the experimental work.

\bibliographystyle{elsarticle-num} 

 % When in Tex format
%\bibliography{library}

\appendix       

    \section{Immobilization capsule geometry}
        \label{sec:appendix_capsule_visualisation}
        
\begin{figure}[htpb!]
            \centering
                \includegraphics[width=0.6\textwidth]{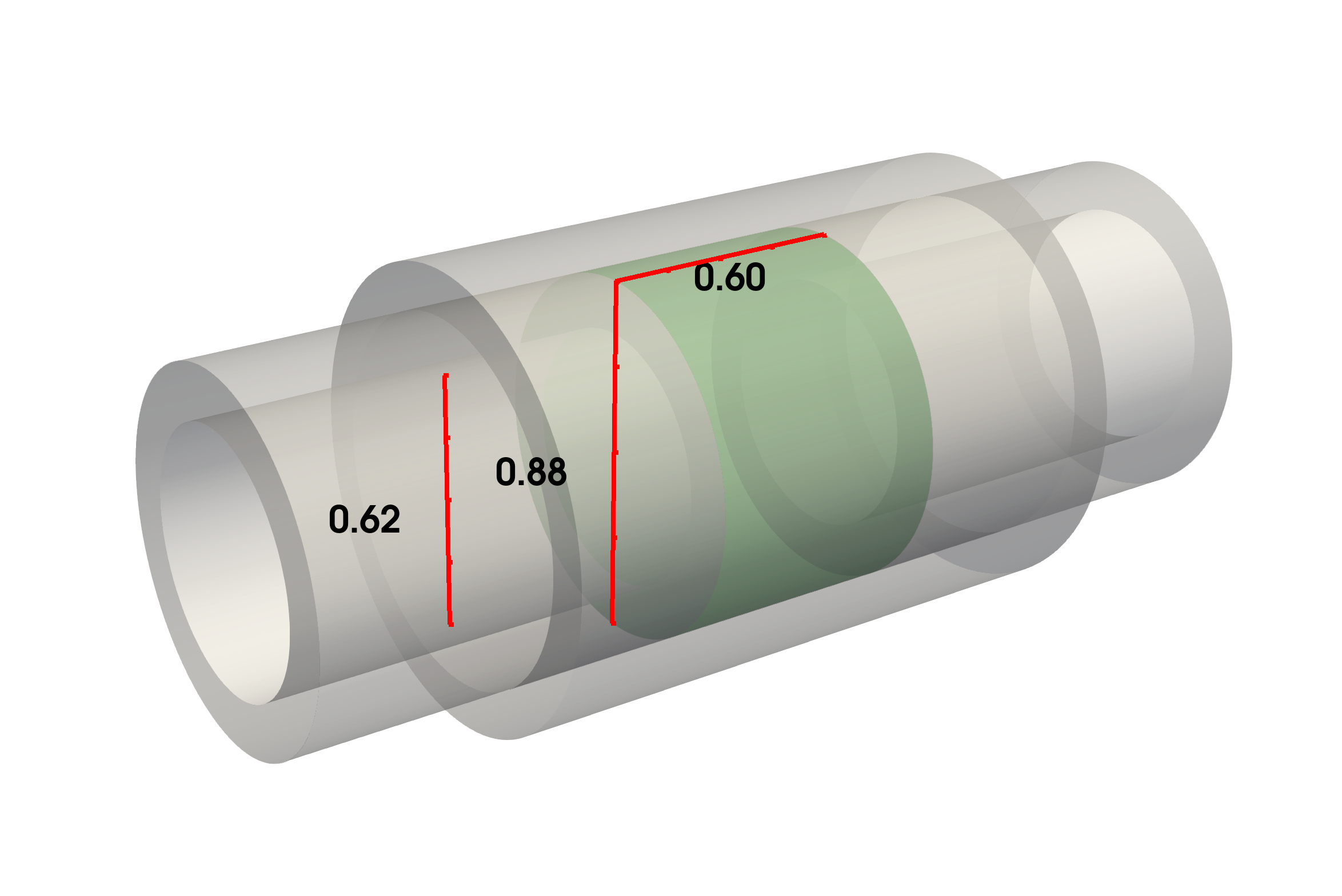}
            \caption{Capsule schematic where monoliths are immobilised. The green region represents a porous monolith. Length units are centimetres.}
            \label{fig:capsule_schema}
        \end{figure}

        \textbf{Figure~\ref{fig:capsule_schema}} shows of the acrylic tube assembly in which each synthesised monolith is immobilised. 

    \section{Porous characteristics distributions}
        \label{sec:appendix_pore_size_distributions}
        
\begin{figure}[htpb!]
            \centering
                \includegraphics[width=\textwidth]{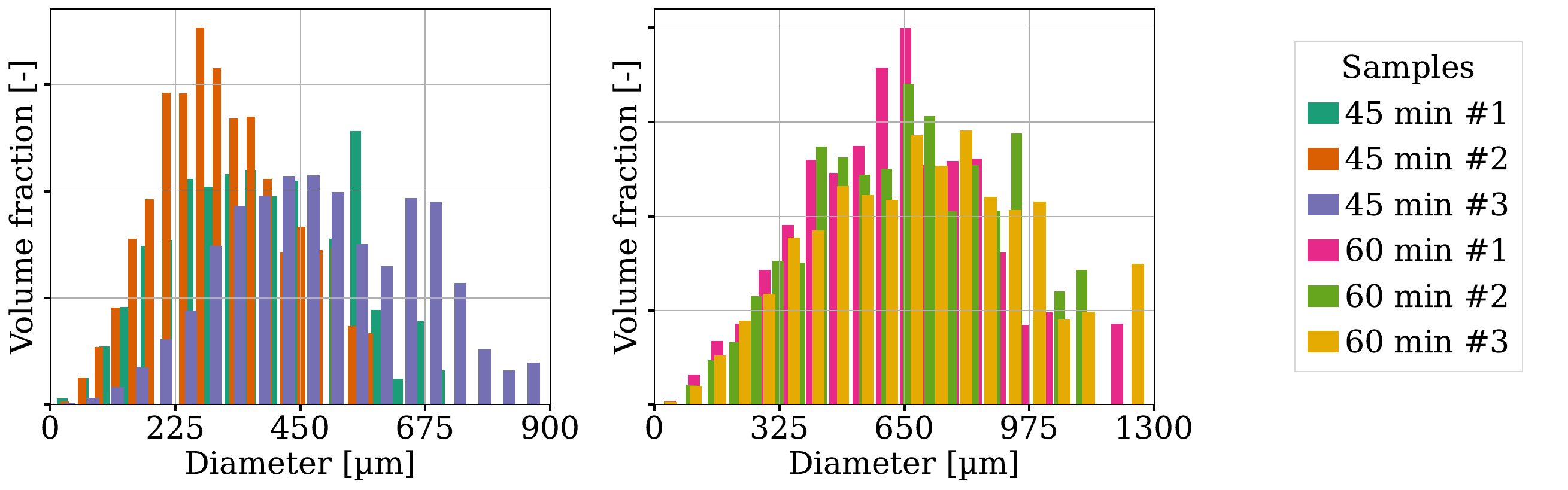}
            \caption{Volume-based pore size distribution of each sample.}
            \label{fig:pore_size_distribution}
        \end{figure}

        \textbf{Figure~\ref{fig:pore_size_distribution}} shows the pore size distribution of each monolith (\SI{45}{\minute} and \SI{60}{\minute} annealing time,~\#$1-3$). The distributions follow Gaussian curves, which confirms the decision to describe the distributions using their mean and standard deviations.
    
\begin{figure}[htpb!]
            \centering
                \includegraphics[width=\textwidth]{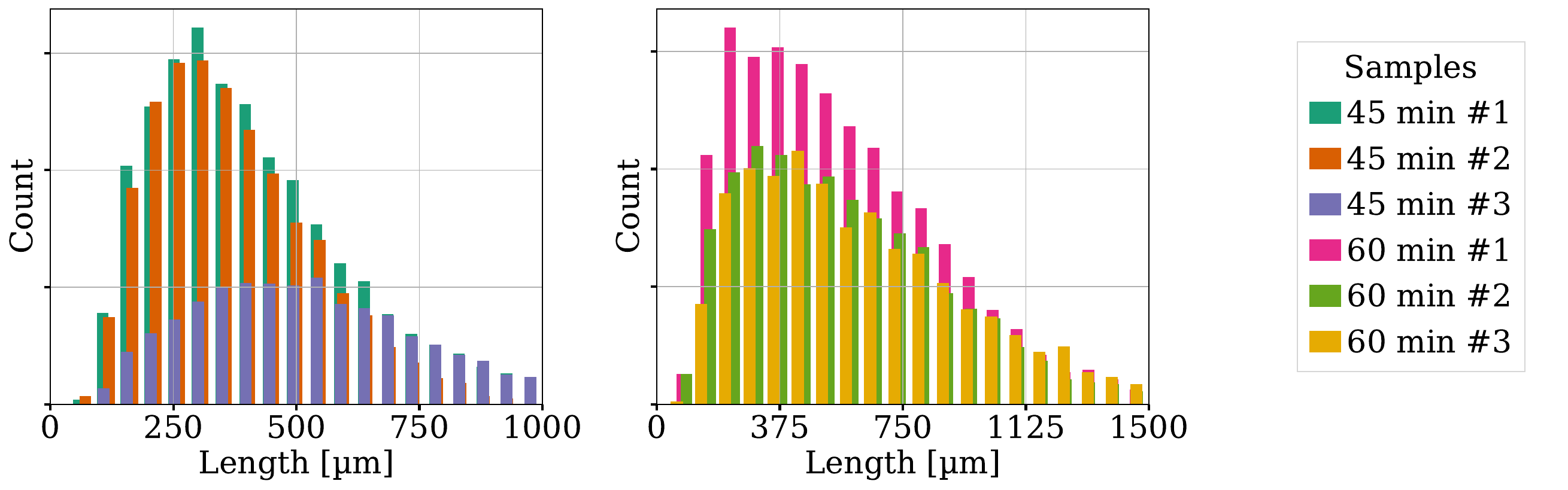}
            \caption{Number-based distance between pairs of connected pores, for each sample.}
            \label{fig:pore_distance_distribution}
        \end{figure}

        \textbf{Figure~\ref{fig:pore_distance_distribution}} shows the distribution of the distance between pairs of connected pores. In this case, the distributions are log-normal.
    
    \section{Transmission electron microscopy of PdNP@silicone} 
        \label{sec:tem_pdnp}
    
\begin{figure}[htpb!]
            \centering
                \includegraphics[width=0.8\textwidth]{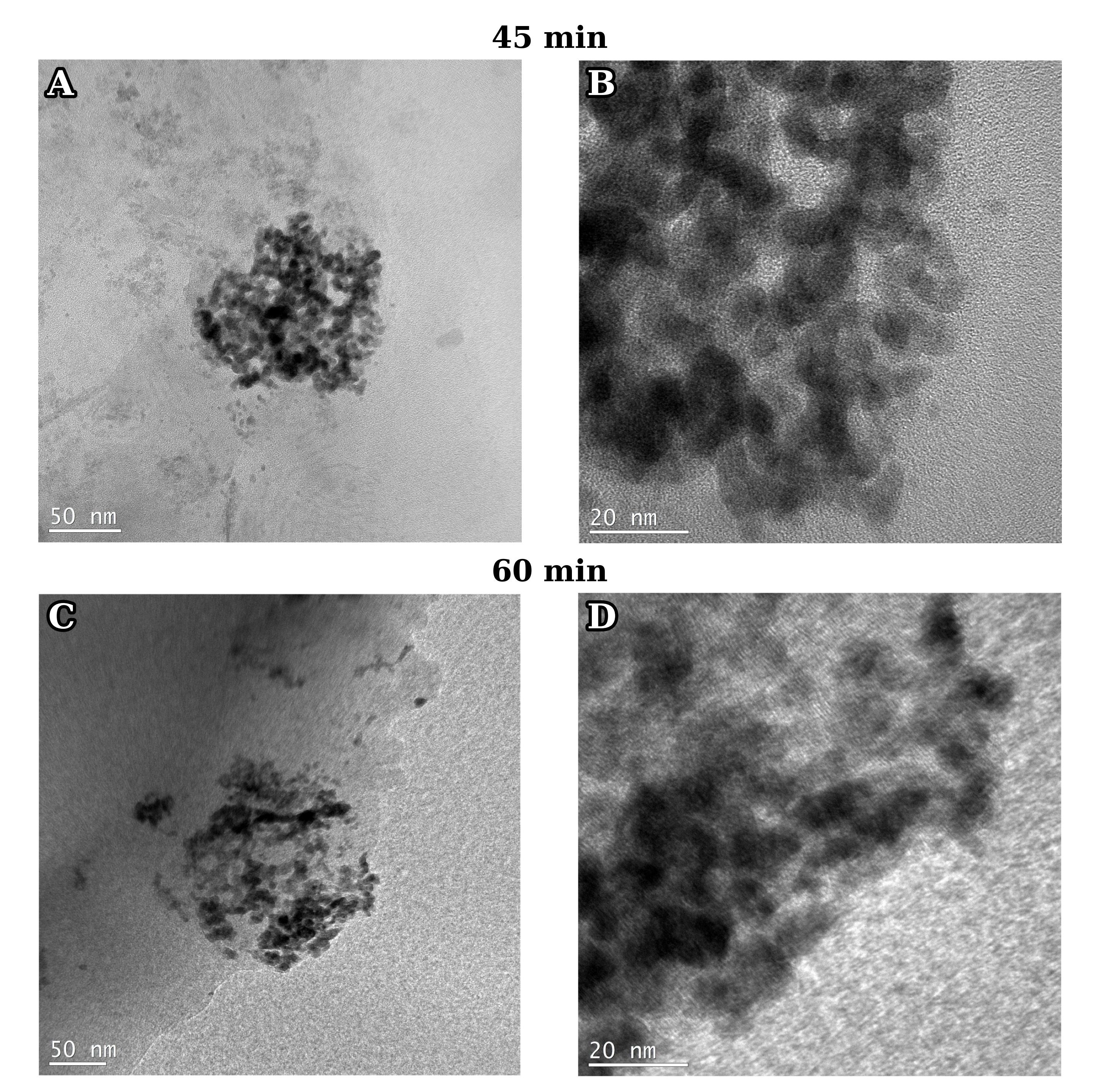}
            \caption{Composite view of captured TEM images of samples at \SI{45}{\minute} and \SI{60}{\minute} annealing durations. A and B show the same cluster of PdNPs on the \SI{45}{\minute} sample at scales of \SI{50}{\nano\metre} and \SI{20}{\nano\metre}, respectively. C and D show another cluster of PdNPs on the \SI{60}{\minute} sample at scales of \SI{50}{\nano\metre} and \SI{20}{\nano\metre}, respectively.}
            \label{fig:tem_global_view}
        \end{figure}

\begin{figure}[htpb!]
            \centering
                \includegraphics[width=1\textwidth]{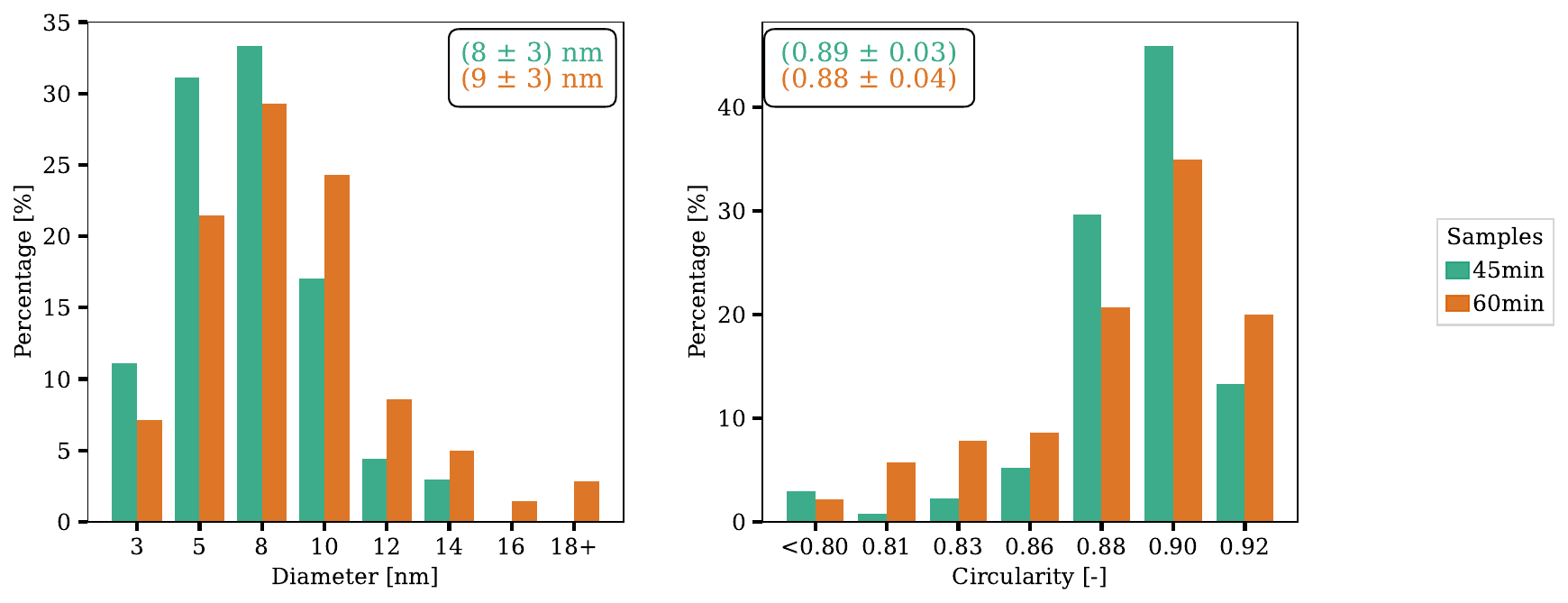}
            \caption{Number-based histogram of particle diameter and circularity based on TEM images of samples of \SI{45}{\minute} and \SI{60}{\minute} annealing durations.}
            \label{fig:histogram_sphericity_dp}
        \end{figure}

        We analysed the size, location and composition of the synthesised PdNPs using Transmission Electron Microscopy~(TEM) and Energy Dispersive Spectroscopy~(EDS) for samples of both annealing times: \SI{45}{\minute} and \SI{60}{\minute}. We analysed images shown in \textbf{Figure~\ref{fig:tem_global_view}} using ImageJ and computed circularity from the computed area and perimeters. \textbf{Figure~\ref{fig:histogram_sphericity_dp}} shows the distributions of each analysed sample, both for particle diameter and circularity. 
        
    \section{Experimental setup for reactive flow experiments}
        \label{sec:experimental_setup}
        
\begin{figure}[htpb!]
            \centering
            \includegraphics[width=1\textwidth]{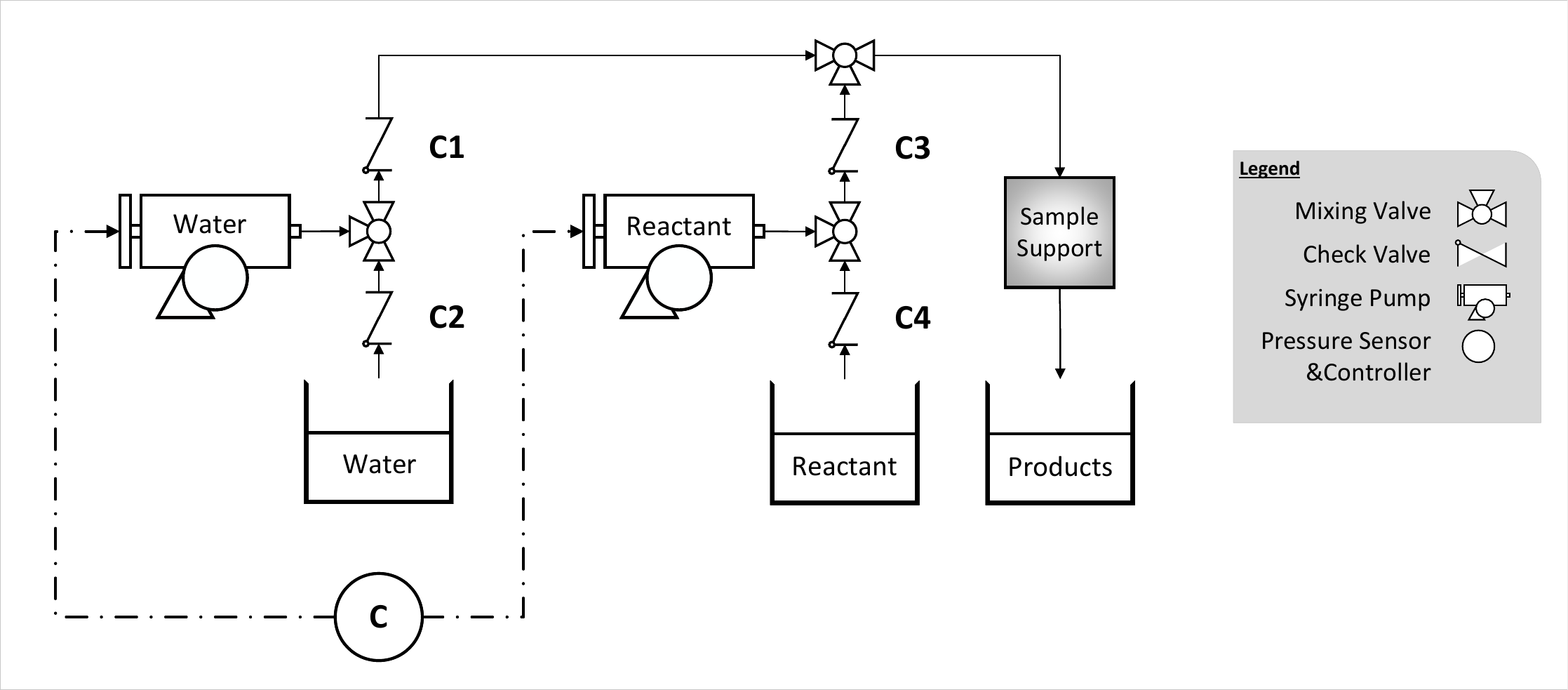}
            \caption{Schematic of the experimental setup used to run reactive flow experiments for the conversion of \textit{p}-nitrophenol catalysed by PdNP@silicone monoliths. The controller~(C) activates the filling and emptying of the water and reactant syringes independently, taking advantage of the variation between positive and negative pressures to open and close check valves~(C1$-$C4).}
            \label{fig:schema_montage}
        \end{figure}

\begin{figure}[htpb!]
        	\centering
        	\includegraphics[width=0.6\textwidth]{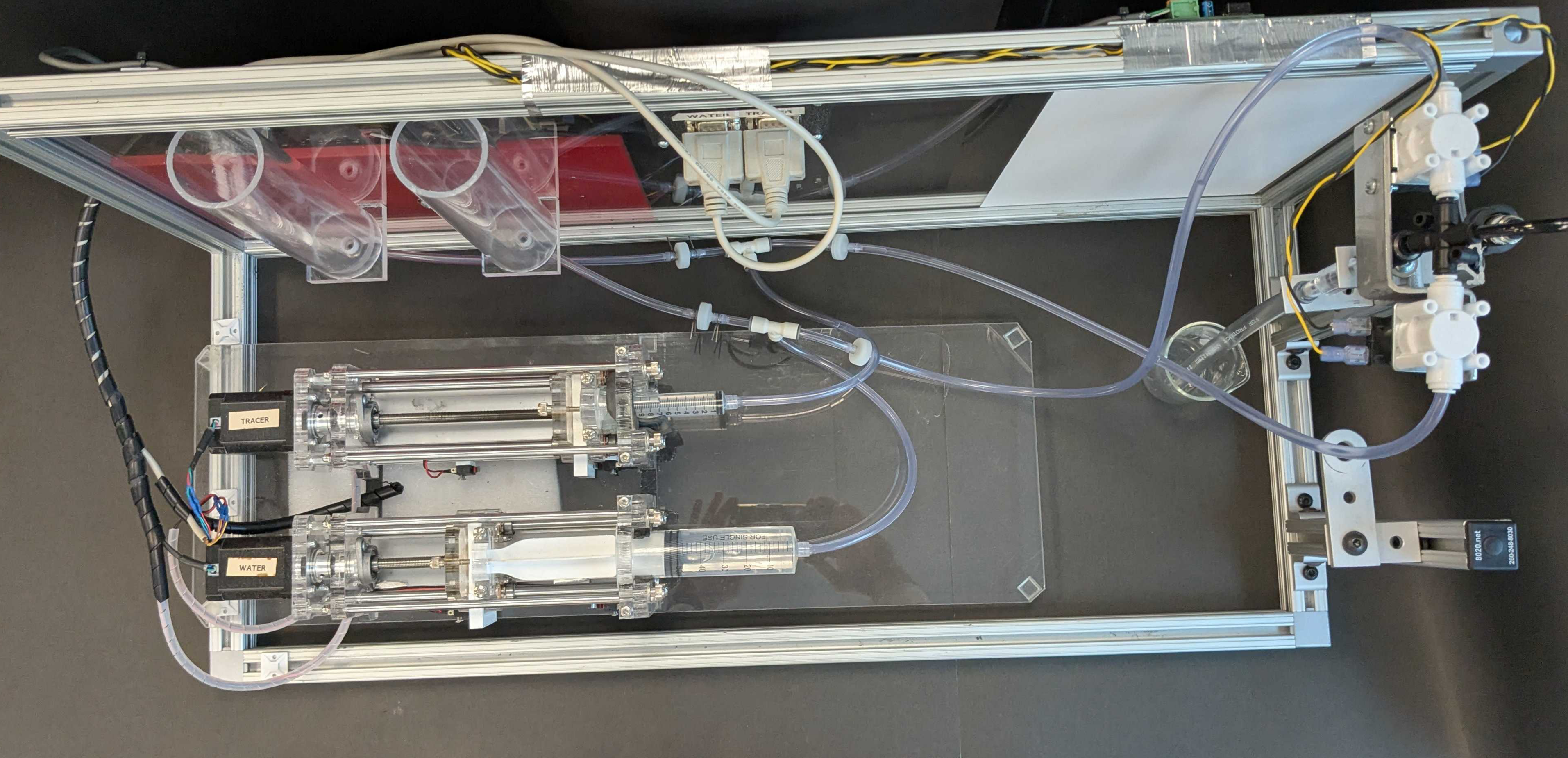}
        	\caption{Photograph (top view) of the experimental setup. The two columns are each connected to their respective syringe and pumps, used for reactant and for distilled water. The check valves allow automatic filling and emptying of the syringe through the samples, located at the right. }
        	\label{fig:photograph_setup}
        \end{figure}

        \textbf{Figures~\ref{fig:schema_montage}} (schematic) and \textbf{\ref{fig:photograph_setup}} (picture) show the setup used for reactive flow experiments. Two syringes and their pumps are used to flush with distilled water and inject reactants according to preset sequences. The filling of the syringes is enabled by check valves that allow to fill syringes from their reservoirs when they are pulled (C2 and C4) while downstream valves are closed (C1 and C3). The emptying of syringes to the porous monolith samples closes the upstream valves (C2 and C4) while opening the downstream valves (C1 and C3). 

     \section{Grid convergence} 
        \label{sec:grid_convergence_annex}
    
        In this section we address grid convergence considerations. 
        %We do not address time step convergence since we are only interested in steady-state metrics. 
        We assume that the flow field is sufficiently resolved in space, based on convergence analyses of previous work studying similar monoliths~\cite{guevremont2024poreresolvedcfddigitaltwin}. We ran grid convergence simulations using the kinetic parameters set in \textbf{Section~\ref{sec:calibration_transferability}} for the sample \SI{60}{\minute}~\#2. 
        Because these are interface-dominated PRCFD simulations with locally refined meshes and steep near-wall concentration gradients, we use $p=1$ as a conservative value for computing grid convergence indices~(GCI) and use a security factor~(FS) of $1.25$~\cite{oberkampf2010vv,roache1998gci}. 
        Grids are refined systematically and span two refinement levels each, with coarser cells in the bulk and finer cells at the interface.
        The results are reported in \textbf{Table~\ref{tab:gci_conversion}}.
        
        \begin{table}[htpb!]
            \centering
            \caption{Grid sensitivity of conversion (fixed order $p=1$) for sample \SI{60}{\minute}~\#2 at coarse~(c), medium~(m) and fine~(f) levels. Base flow rate $Q=$~\SI{0.013}{\milli\litre\per\second}. Discretization errors are $\varepsilon_{f,m}=|X_f-X_m|$ and $\varepsilon_{m,c}=|X_m-X_c|$. }
            \label{tab:gci_conversion}
            \begin{tabular}{c c c}
                \toprule
                \cmidrule(lr){2-3}
                \textbf{Flow rate}& Q & 10Q \\
                \midrule
                \multicolumn{1}{c}{\textbf{Cell size} [\textmu m]} & \multicolumn{2}{c}{\textbf{Conversion} $X$} \\
                \cmidrule(lr){1-3}
                46.88~(c) & 0.1586 & 0.0326 \\
                23.44~(m) & 0.1599 & 0.0317 \\
                11.72~(f) & 0.1579 & 0.0279 \\
                \cmidrule(lr){1-3}
                \textbf{$\varepsilon_{\mathrm{m,c}}=|X_\mathrm{m}-X_\mathrm{c}|$} & 0.0013 & 0.0009 \\
                \textbf{$\varepsilon_{\mathrm{f,m}}=|X_\mathrm{f}-X_\mathrm{m}|$} & 0.0020 & 0.0037 \\
                \cmidrule(lr){1-3}
                \textbf{Richardson extrapolation} & 0.1559 & 0.0242 \\
                \cmidrule(lr){1-3}
                \multicolumn{1}{c}{\textbf{Cell size} [\textmu m]} & \multicolumn{2}{c}{\textbf{GCI} [\%]} \\
                \cmidrule(lr){1-3}
                46.88~(c) & - & - \\
                23.44~(m) & 1.02 & 3.69 \\
                11.72~(f) & 1.58 & 16.59 \\
                \bottomrule
            \end{tabular}
        \end{table}
        
        Conversion varies little across refinements. The computed GCI values are comparable to other sources of uncertainty (e.g., experimental variability and kinetic calibration), particularly at the base flow rate $Q$. At $10Q$, GCI increases from the medium to the fine level, indicating that this refinement is outside the asymptotic regime. Because the medium-to-fine correction exceeds the coarse-to-medium correction, Richardson extrapolation should be interpreted qualitatively rather than as a strict error bound.
        The GCI is used here as a practical measure of grid sensitivity and uncertainty indicator, rather than as a rigorous asymptotic error estimate. 

    \section{Composite solid definition}
        \label{sec:composite_solid_definition}

        We define the solid used for the reactive flow simulations from a composite. The shapes used are:
        
        \begin{enumerate}
            \item Monolith, bounded by a cylinder of radius $0.44$ and length $0.6$, centred around $(0,0,0)$ and oriented in the X-axis;
            \item Hyperrectangle, centred around $(0,0,0)$, with lengths of $(1.2,1.2,1.2)$;
            \item Cylinder of radius $0.44$ and length $0.6$, centred around $(0,0,0)$ and oriented in the X-axis;
            \item Cylinder of radius $0.31$ and length $1.2$, centred around $(0,0,0)$ and oriented in the X-axis.
        \end{enumerate}
        
        The boolean operations are:
        \begin{enumerate}
            \setcounter{enumi}{4}
            \item Hollowed hyperrectangle $ = 2 \setminus 3$;
            \item Pierced hollowed hyperrectangle $ = 5 \setminus 4$;
            \item Support with inserted monolith $ = 6 \cup 1$.
        \end{enumerate}

    \section{Calibration data per flow rate} 
        \label{sec:calibration_data_per_flow_rate}
    
\begin{figure}[htpb!]
            \centering
                \includegraphics[width=0.8\textwidth]{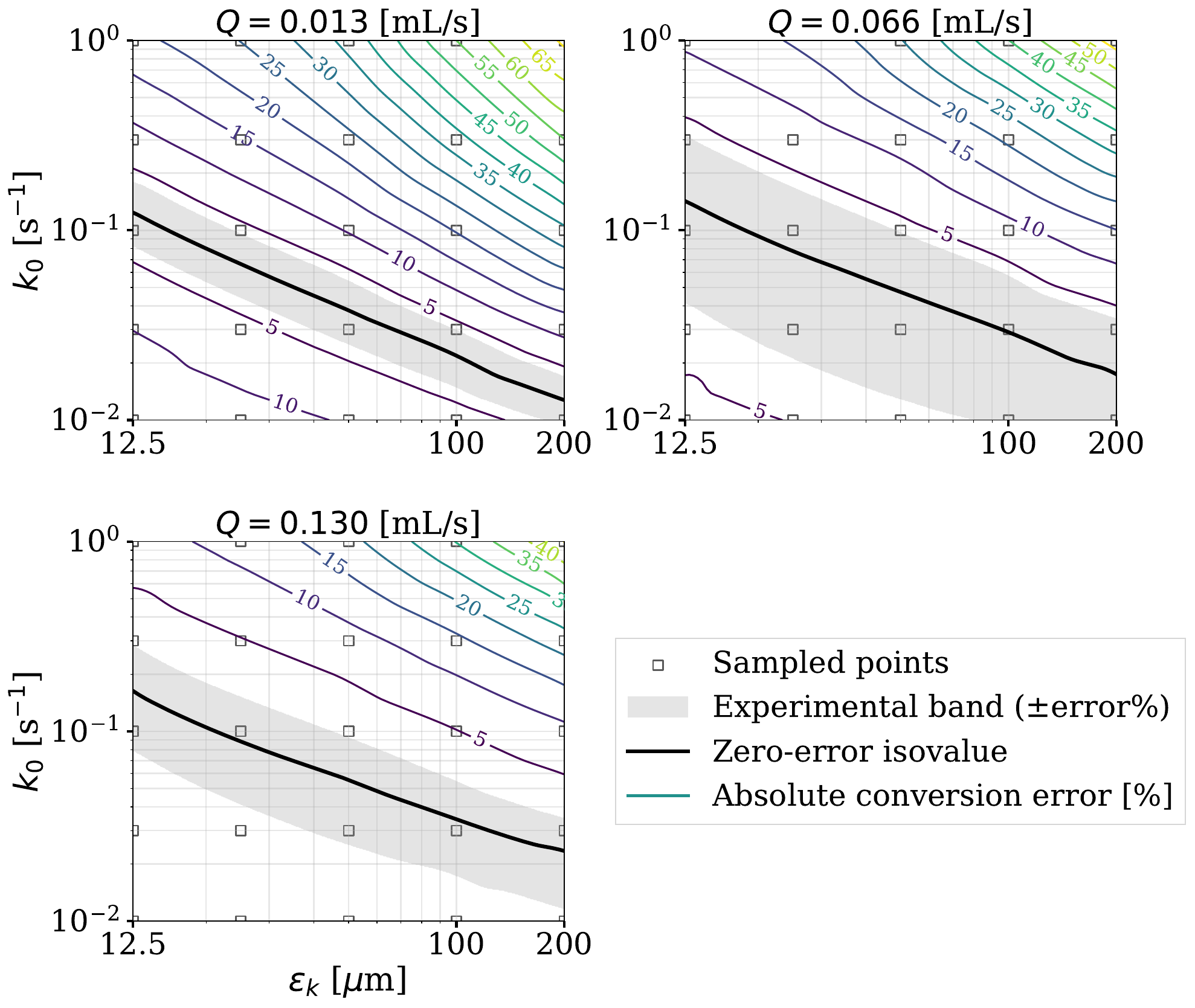}
            \caption{Simulated absolute conversion error for various reaction constants $k_0$ and Gaussian distribution thickness $\varepsilon_k$, for different flow rates, for sample \SI{60}{\minute}~\#2. The curves are obtained from interpolation.}
            \label{fig:calibration_per_flow_rate}
        \end{figure}

        \textbf{Figure~\ref{fig:calibration_per_flow_rate}} shows each flow rate's individual calibration map for the sample \SI{60}{\minute}~\#2. The subplots present different optimal curves (zero-error isovalue), but all follow the same trends. The isovalues being more concentrated in lower flow rates make sense given that accessibility to the surface is higher for higher residence times.

    \section{Visual comparison of 45 min samples}
        \label{sec:comparison_45min_samples}

\begin{figure}[htpb!]
            \centering
                \includegraphics[width=0.45\textwidth]{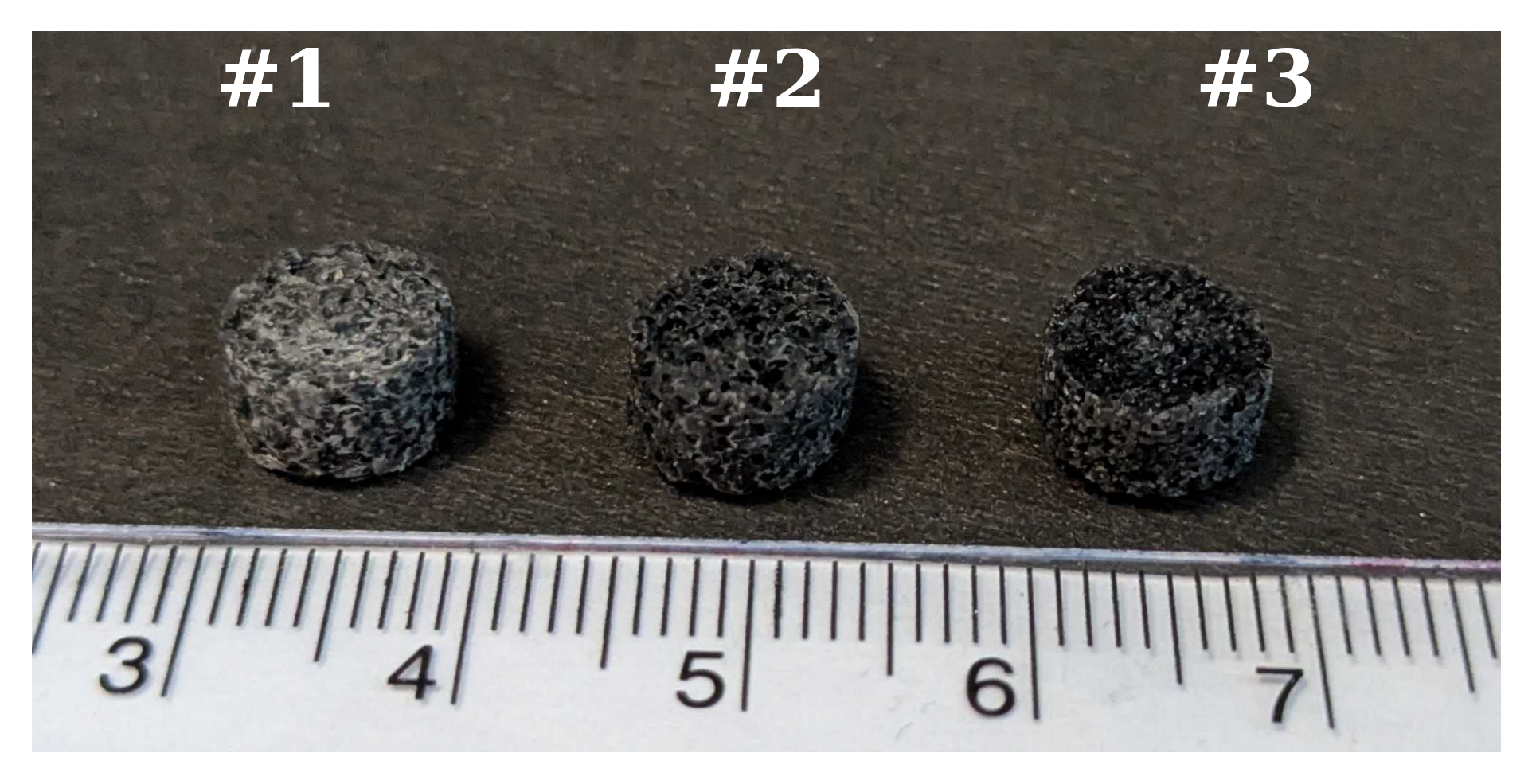}
            \caption{Samples of \SI{45}{\minute}~\#1$-$3 annealing durations after PdNPs synthesis. }
            \label{fig:three_45mins}
        \end{figure}

        \textbf{Figure~\ref{fig:three_45mins}} shows the \SI{45}{\minute}~\#1$-$3 samples. The colour difference for \#1 is consistent with the presence of a contaminant layer, likely a polymer film formed during synthesis.

    \section{Batch reaction results}
        \label{sec:batch_reaction_results}

\begin{figure}[htpb!]
            \centering
                \includegraphics[width=0.4\textwidth]{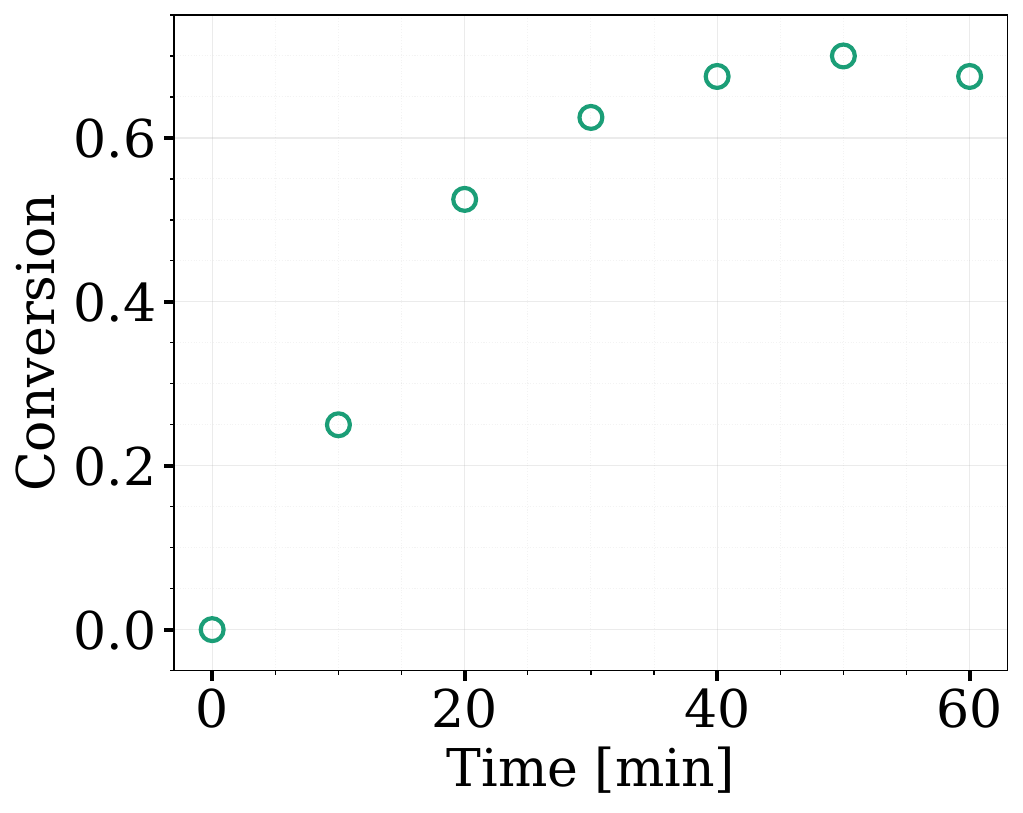}
            \caption{Conversion of \textit{p}-nitrophenol in a batch reaction with excess NaBH$_4$ when in contact with sample \SI{60}{\minute}~\#3, over a duration of \SI{60}{\minute}. }
            \label{fig:reactivity_no_flux}
        \end{figure}

        \textbf{Figure~\ref{fig:reactivity_no_flux}} shows the conversion of \textit{p}-nitrophenol, when placed in a glass vial for a duration of \SI{60}{\minute} with the sample \SI{60}{\minute}~\#3. 

    \section{Generated structured media} 
        \label{sec:generated_structured_media}
    
\begin{figure}[htpb!]
            \centering
            \includegraphics[width=\textwidth]{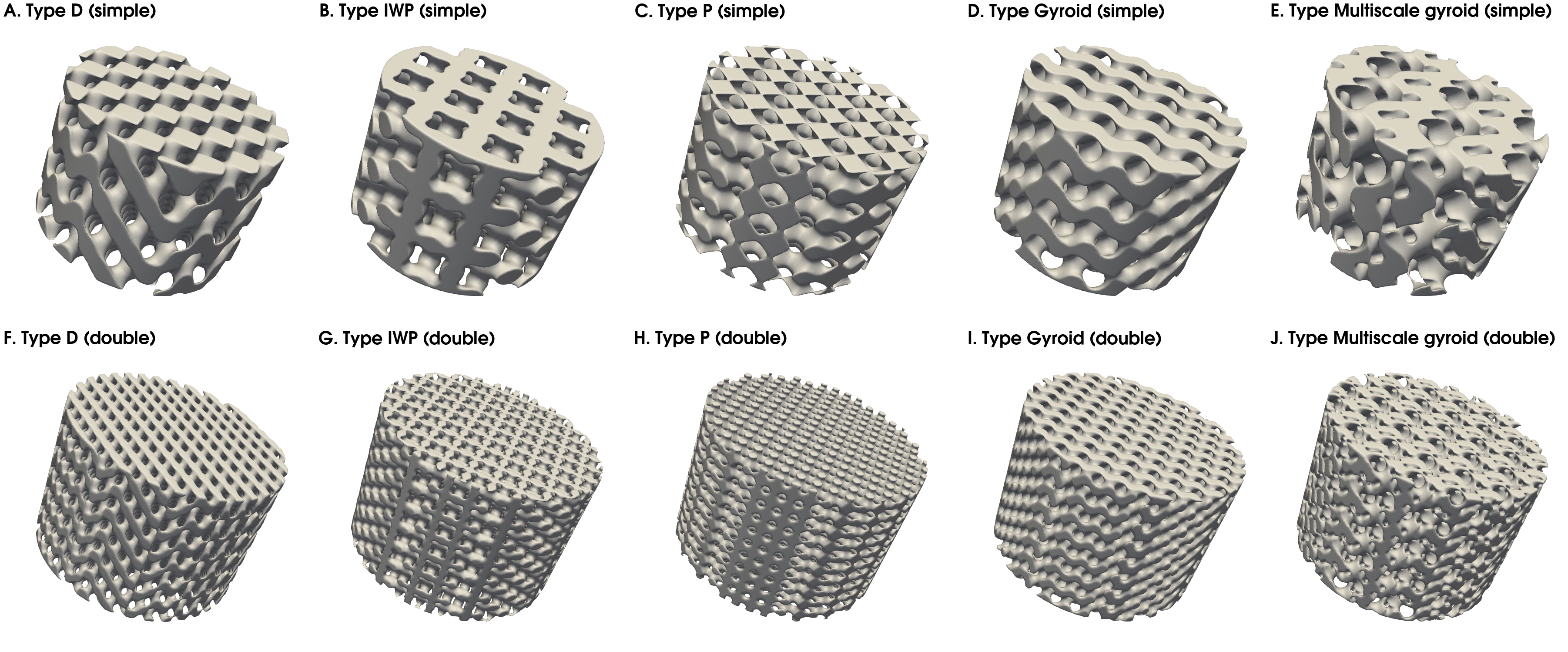}
            \caption{Structured media generated using Cesogen.
            }
            \label{fig:visualisation_10_tpms}
        \end{figure}

        \textbf{Figure~\ref{fig:visualisation_10_tpms}} shows the generated structured media used for PRCFD simulations in \textbf{Section~\ref{sec:structural_design}}.
        
    \section{Surface efficiency cumulative distribution function} 
        \label{sec:surface_distribution_function}
    
\begin{figure}[htpb!]
            \centering
            \includegraphics[width=\textwidth]{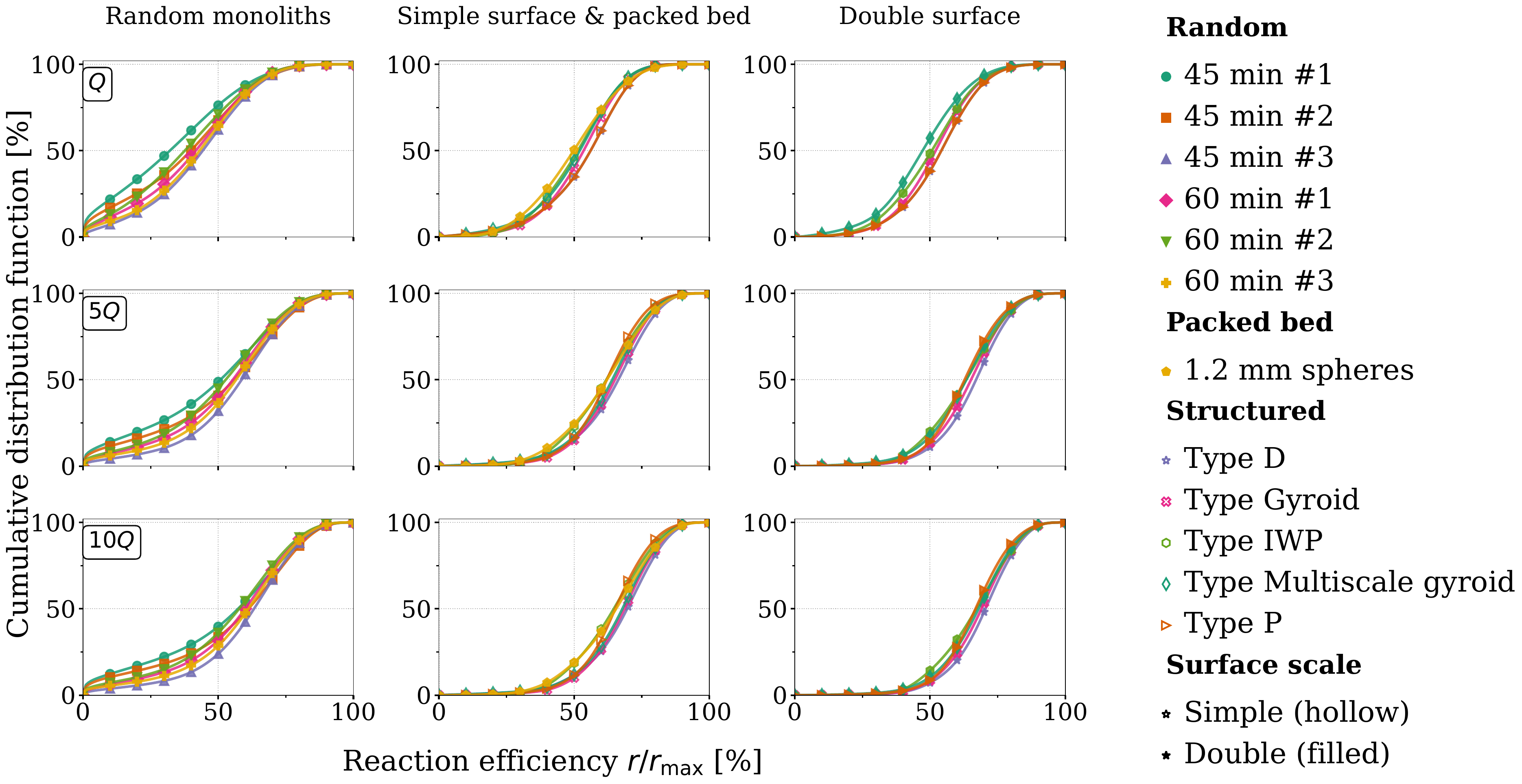}
            \caption{Surface cumulative distribution of the reaction efficiency across digitised random monoliths (\SI{45}{\minute}~\#$1-3$, \SI{60}{\minute}~\#$1-3$), structured TPMS-based monoliths (simple and double surface types D, Gyroid, IWP, Multiscale gyroid, P), and a packed bed.
            Each row corresponds to a flow rate, where the base flow rate $Q=0.013$~\SI{}{\milli\litre\per\second}. 
            The first column displays the random monoliths, the second the simple surface TPMS-based structures and the packed bed, and the third the double surface TPMS-based structures.
            }
            \label{fig:surface_efficiency_pdf}
        \end{figure}

        \textbf{Figure~\ref{fig:surface_efficiency_pdf}} shows the surface cumulative distribution function for each generated or digitised geometry analysed in \textbf{Section~\ref{sec:structural_design}}. For the random monoliths, the high portion of the surface at lower reaction efficiency supports the diagnosis that a significant portion of the catalysis surface is not accessed by the reactants, although there are additional sharper increases close to $50\%-75\%$ reaction efficiency.
        
\end{document}